\documentclass[aps,prev,onecolumn,preprintnumbers,floatfix,nofootinbib]{revtex4-1}
\pdfoutput=1
\usepackage{geometry,amsmath,amsfonts}
\usepackage{slashed}
\usepackage{epsfig}
\usepackage{latexsym}
\usepackage{graphicx}
\usepackage[justification=RaggedRight]{caption}
\usepackage{amssymb}
\usepackage{subfig}
\usepackage{color}
\usepackage{multirow}
\usepackage{color}
\usepackage{comment}
\usepackage{rotating}
\usepackage{ifthen}
\usepackage{epsfig}
\usepackage{xspace}

\definecolor{orange}{rgb}{1,0.4,0}
\definecolor{green}{rgb}{0,0.65,0}
\definecolor{rossos}{rgb}{0.8,0.2,0.3}
\definecolor{bluscuro}{rgb}{0.15, 0.2, .85}
\definecolor{bluchiaro}{cmyk}{1,.3,0.,0.1}

\makeatother   

\newcommand{\GeV}{{\rm \,GeV}}
\newcommand{\TeV}{{\rm \,TeV}}
\newcommand{\MeV}{{\rm \,MeV}}

\newcommand{\HiTo}{\texttt{HiggsTools}\xspace}
\newcommand{\HiBo}{\texttt{HiggsBounds}\xspace}
\newcommand{\HiSi}{\texttt{HiggsSignals}\xspace}

\newcommand{\met}{\slashed{E}_T}

\newcommand{\sab}{\sin\left(\alpha-\beta\right)}
\newcommand{\cab}{\cos\left(\alpha-\beta\right)}

\begin{document}

 \title{Is there a (Pseudo)Scalar at $95\GeV$?}

\author{Giorgio Arcadi$^{a,b}$}
\email{{giorgio.arcadi@unime.it}}
\author{Giorgio Busoni$^{c,d}$}
\email{giorgio.busoni@adelaide.edu.au}
\author{David Cabo-Almeida$^{a,b}$}
\email{david.cabo@ct.infn.it}
\author{Navneet Krishnan$^c$}
\email{navneet.krishnan@anu.edu.au}
\vspace{0.1cm}
\affiliation{${}^a$ Dipartimento di Scienze Matematiche e Informatiche, Scienze Fisiche e Scienze della Terra, \\ Universita degli Studi di Messina, Viale Ferdinando Stagno d'Alcontres 31, I-98166 Messina, Italy}
\affiliation{${}^b$ INFN Sezione di Catania, Via Santa Sofia 64, I-95123 Catania, Italy}
\affiliation{${}^c$ ARC Centre of Excellence for Dark Matter Particle Physics, \\
Department of Fundamental and Theoretical Physics, Research School of Physics,\\
The Australian National University, Canberra ACT 2601, Australia}
\affiliation{${}^d$ ARC Centre of Excellence for Dark Matter Particle Physics, \\
University of Adelaide, South Australia 5005, Australia}

\begin{abstract}

We discuss the possibility of interpreting the recent experimental hints in favour of a 95 GeV resonance with extensions of the Standard Model featuring an extra Higgs doublet and SM scalar (2HDM+s) or pseudoscalar singlet (2HDM+a). We compare the possibility of reproducing the experimental anomalies with the theoretical constraints on the extended Higgs sector as well as complementary bounds coming from flavour physics and other colliders searchers. For both the 2HDM+s and 2HDM+a we will consider a generic natural flavour conserving (NFC) model as well as the customary Type-I, -II, -X and -Y configurations of the Yukawa coupling to the BSM Higgs bosons. We find that no model is able to reproduce all experimental hints, and plot the $2\sigma$ C.L. favoured regions in the parameter space.

\end{abstract}

\maketitle

\section{Introduction}

The Standard Model of Particle Physics (SM) has had incredible success in explaining physical phenomena with high precision in the last decades. Formulated in the mid-seventies, its last missing piece, the Higgs boson (which we denote in this paper as $h_{125}$), was ultimately discovered in 2012 at the Large Hadron Collider (LHC) \citep{CMS:2012qbp,ATLAS:2012yve} with a mass of $m_{h_{125}}=(125.11 \pm0.11 \GeV)$ \citep{ATLAS-CONF-2023-037}. We know, however, that the SM cannot be the ultimate theory of nature, for reasons both theoretical and experimental. Three of these are the evidence of neutrino oscillations, requiring neutrinos to have mass \citep{Super-Kamiokande:1998kpq}, that it gives no explanation for the matter-antimatter asymmetry of the universe \citep{Kuzmin:1985mm}, and it does not provide a Dark Matter (DM) candidate \citep{Jungman:1995df,Bertone:2016nfn}. 
\\
\\
Since the discovery of the Higgs boson, physicists have measured its properties to understand if it is the only fundamental scalar particle, or if it is part of a Beyond the Standard Model (BSM) theory incorporating an extended scalar sector. 
Indeed, many BSM theories predict or incorporate extended scalar sectors, the most common being theories including additional Higgs doublets and/or scalar singlets. For example, the existence of two Higgs doublets is a requirement of the MSSM \citep{Fayet:1976cr,Nilles:1983ge,Barbieri:1987xf,Martin:1997ns}, while the NMSSM incorporates two doublets and an additional singlet \citep{Ellis:1988er,Maniatis:2009re}. Extended scalar sectors have attractive features, such as being able to generate a strong first-order phase transition to allow Electroweak (EW) baryogenesis  that cannot be achieved in the SM \citep{Morrissey:2012db}, or providing simple extensions of the SM where one can generate neutrino masses and have Dark Matter candidates, such as in the scotogenic models \citep{Tao:1996vb,Ma:2006km,Hirsch:2013ola}.
\\
\\
In the presence of an extended scalar sector we expect a deviation in the value of the couplings of the 125 GeV Higgs boson $h_{125}$ from the predictions of the SM due to the mass mixing of the latter with the additional scalars in the theory. Extended studies at the LHC strongly disfavor sizable mass mixing, and so it is customary to assume the so called alignment limit: a mathematical condition automatically enforcing the couplings of $h_{125}$ to be SM-like. Notice that, while the couplings of the $h_{125}$ Higgs to gauge bosons and third generation fermions have been constrained to be SM-like \citep{CMS:2018uag,ATLAS:2022vkf}, constraints on the Higgs self-couplings are still far from being able to restrict their allowed values close to the ones expected by the SM \citep{CMS:2022dwd}. 
\\
\\
\noindent The most popular extension of the Higgs sector features the addition of a second doublet, hence realizing a two Higgs doublet model (2HDM). After Electroweak Symmetry Breaking (EWSB), both doublets gain vevs, generating a physical mass spectrum including, besides $h_{125}$, two additional neutral scalars, one CP-even and one CP-odd\footnote{It is implicitly assumed here that the scalar sector preserves CP.}, and a charged state. Currently, depending on its mass $m_\phi$, a light (pseudo)scalar faces the following experimental constraints and detection challenges:

\begin{itemize}
    \item The $h_{125}$ Higgs must have branching ratios (BR) very similar to those predicted by the SM. If $2m_b<m_\phi<m_h/2\sim 62\GeV$, unless the $h\phi\phi$ coupling is negligibly small for some reason, one would get a considerable BR $h_{125}\rightarrow \phi\phi$,  with $\phi$ decaying mostly to $b\bar{b}$ and $\tau^+\tau^-$. The observed decays of the $h_{125}$ therefore place strong constraints on the existence of additional scalars in this mass range. Additionally, $h_{125}$ decays through intermediate scalars have been explicitly searched for \citep{CMS:2022xxa}, with null results.
    
    \item The previous mass range can be extended to slightly larger masses, namely up to around 75 GeV, by considering the additional 3 body decay $h_{125}\rightarrow \phi\phi^*$. As $m_\phi$ grows, this decay channel becomes more and more off-shell and therefore small. While the exact limits depend on the $h\phi\phi$ coupling, overall we can expect the mass range $10\GeV<m_\phi\lesssim75\GeV$ to be tightly constrained, both for scalars and pseudoscalars.
    
    \item  On the other hand, if $m_\phi>2m_t\sim 340\GeV$, given that the additional scalars' couplings to SM fermions are expected to be proportional to the fermion masses, strong constraints from $t\bar{t}$ resonance searches would apply. Again, this limit holds both for scalars and pseudoscalars.
    
    \item Away from the alignment limit, an additional pseudoscalar particle with  $m_\phi>m_{h_{125}}+m_Z\sim210\GeV$ would be constrained by the $\phi\rightarrow Zh_{125}$ decay, and an additional scalar particle with $m_\phi> 2m_W\sim 160\GeV$ would be constrained by the $\phi\rightarrow WW$ decay. These have been searched for with null results in the mass ranges $m_\phi>225\GeV$ and $m_\phi>200\GeV$ respectively \citep{CMS:2015uzk,CMS:2019qcx,ATLAS:2017uhp,ATLAS:2018sbw,ATLAS:2022eap}.
    
    \item However, the remaining mass ranges, $75\GeV\lesssim m_\phi<210\GeV$ for a pseudoscalar and $70\GeV<m_\phi<160\GeV$ for a scalar, and in particular the mass window $75\GeV\lesssim m_\phi\lesssim 100\GeV$, are very difficult to probe. 
    In the latter mass range, off shell $\phi\rightarrow W W^\star$ is highly kinematically suppressed (and is completely absent for a pseudoscalar), and the new scalar usually decays mainly to $b\bar{b}$. Due to the very large background at LHC, pure $b\bar{b}$ resonance searches usually consider a much higher mass window \citep{CMS:2022eud}, and one can search for $\phi\rightarrow b\bar{b}$ only when produced in association with some other particle, for example an EW boson.
    Additional searches that could detect a light scalar in this mass range are searches for di-tau resonances $\phi\rightarrow \tau^+\tau^-$ and diphoton $\phi\rightarrow\gamma\gamma$ searches. These searches may suffer from a larger than usual Drell-Yan background in this mass range, especially the range $m_\phi\in[85,95]\GeV$ very close to the $Z$ mass.
\end{itemize}

\noindent Interestingly, there have recently been mounting hints for a new light (pseudo)scalar of mass $m_\phi\sim95\GeV$ exactly from these channels:

\begin{itemize}
    \item In a search for additional Higgs bosons $\phi$ in $\tau\tau$ final states using the full Run 2 dataset by CMS \citep{CMS:2022goy}, a $3.1\sigma$ local excess of events was found assuming gluon fusion production $gg\rightarrow\phi$ at an invariant mass $m_{\tau\tau}\sim100\GeV$. No corresponding excess is found in the $gg\rightarrow \phi b\bar{b}$ channel. Figure 11 of \citep{CMS:2022goy} indicates that for a scalar with mass $\sim 100\GeV$, the results are consistent with no events of production in association with $b$ quarks, and non-zero events with no $b$-tagged jets. The signal at $m_{\tau\tau}\sim95\GeV$ has a $2.6\sigma$ local excess that is quantified as $\sigma\times BR = 7.8_{-3.1}^{+3.9}pb$. This, normalised to the gluon-fusion production cross section of a $95\GeV$ boson \citep{LHCHiggsCrossSectionWorkingGroup:2016ypw} times its SM branching ratio, means a signal strength $\mu_{\tau\tau}=1.22_{-0.48}^{+0.62}$. 
    
    \item Recently, an analysis from CMS of the full Run 2 dataset with $138 fb^{-1}$ \citep{CMS:2023yay} reported a $2.8\sigma$ local excess in the diphoton invariant-mass distribution at $m_{\gamma\gamma} = 95.4 \GeV$. However, this excess is mainly due to 2016 data, without a significant excess appearing in 2017 data, and only a small excess in 2018 data, as shown in Fig. 7 of \citep{CMS:2023yay}. The analysis considers all production modes (gluon fusion, Vector Boson Fusion (VBF), Vector Boson (VH) associated production, and $t\bar{t}$ associated production) and presents results for the combined production modes as well as the separate combinations of gluon fusion plus $t\bar{t}$ and VBF+VH, and for the VBF and VH channels separately. 
    Considering only the analysis with the combined production modes, the resulting signal strength is $\mu_{\gamma \gamma}= 0.33_{-0.12}^{+0.19}$.

    \item A previous CMS search in the diphoton channel with the Run 1 data at $\sqrt{s}=8 \text{ TeV}$ \citep{CMS:2015ocq} also reported an excess at the same mass value of about $2\sigma$. This search was combined with the initial part of the Run 2 dataset \citep{CMS:2018cyk}, with a combined claimed signal strength of $\mu_{\gamma\gamma}=0.6\pm0.2$.
    
    \item Another search for diphoton resonances in the low mass range  $66-110 \GeV$ has recently been presented by ATLAS, using $140 fb^{-1}$ of $13 \text{ } TeV$ $pp$ collisions (full Run 2 dataset) \citep{ATLAS:2023jzc}. This search makes both a model-independent analysis targeting only gluon fusion production modes, and a model-dependent analysis including all SM-like Higgs production modes. While this search does not show any excess above $2\sigma$ for either analysis strategy, they both have a very small excess around $95 \GeV$, with a local significance of $1.1\sigma$ for the model-independent analysis and a significance of $1.7\sigma$ for the model-dependent one. The reported excess converts to a signal strength $\mu_{\gamma\gamma}=0.18_{-0.10}^{+0.10}$ \cite{Biekotter:2023oen}. Combining the data from both CMS and ATLAS in the diphoton channel yields a signal strength $\mu_{\gamma\gamma} = 0.24_{-0.08}^{+0.09}$ \cite{CMS:2018cyk}.
    
    \item In the same mass range, the statistical combination of the final results from the Large Electron Positron (LEP) experiments ALEPH, DELPHI, L3 and OPAL published in 2006 shows an excess in the $e^+e^-\rightarrow Z\phi, \phi\rightarrow bb$ channel at $m_{bb}=98\GeV$ with a local significance of $2.3\sigma$ \citep{LEPWorkingGroupforHiggsbosonsearches:2003ing,Azatov:2012bz,Cao:2016uwt}. Due to the limited dijet mass resolution at LEP this excess could have the same origin as the aforementioned $\gamma\gamma$ and $\tau^+\tau^-$ anomalies. The resulting signal strength is $\mu_{bb}=0.117\pm0.06$.
    
    \item Interestingly, there are two resonance searches with cascade decays $X\rightarrow Yh_{125}$, one from CMS \citep{CMS:2022suh} and one from ATLAS \citep{ATLAS:2023azi}, both assuming only gluon fusion production for the heavy scalar $X$, that also hint towards a new scalar with a mass compatible with $95\GeV$. The CMS search \citep{CMS:2022suh} has the highest global significance for $m_X=1.6\TeV$ and $m_Y=90\GeV$, while the ATLAS search \citep{ATLAS:2023azi} has the lowest p-value for the background-only hypothesis in the bin where $m_X\in[3608,3805]\GeV$ and $m_Y\in[75.7-95.5]\GeV$.
\end{itemize}

\noindent These anomalies have attracted a lot of attention \citep{Belanger:2012tt,Becirevic:2015fmu,Cao:2016uwt,Fox:2017uwr,Haisch:2017gql,Heinemeyer:2021msz,Azevedo:2023zkg,Biekotter:2023jld,Biekotter:2023oen,Dutta:2023cig,Aguilar-Saavedra:2023vpd,Borah:2023, Iguro:2022dok,Iguro:2022fel,Ashanujjaman:2023etj,Cao95GeV,Cao:2019ofo,Bhattacharya:2023lmu}, and several authors \citep{Becirevic:2015fmu,Fox:2017uwr,Haisch:2017gql,Azevedo:2023zkg, Iguro:2022fel}  have tried to interpret (some of) these anomalies in the context of a 2HDM. However, the interpretation in terms of a 2HDM usually generates tension as a 2HDM must simultaneously satisfy three requirements. First, it must reproduce observed signal strengths. This prevents $\tan\beta = \frac{v_2}{v_1}$, where $v_i$ are the vevs of the two doublets, from being too large, suppressing the gluon fusion production cross-section. Second, it must satisfy perturbativity and unitarity, which provide upper bounds on the mass differences between the charged Higgs and the new neutral scalars. Finally, it must satisfy flavour constraints, which do not allow light charged scalars at low $\tan\beta$.
\\
\\

\noindent To address these requirements, the next step in complexity is the 2HDM with the addition of a singlet real scalar (2HDM+s \citep{Bell:2016ekl}) or pseudoscalar (2HDM+a \citep{Ipek:2014gua}).
These two models are popular theoretical benchmarks as they provide a very broad range of collider signatures while being, at the same time, gauge invariant extensions of the Standard Model \cite{LHCDarkMatterWorkingGroup:2018ufk}. Furthermore, these models allow us to consistently, from a theoretical perspective, embed the coupling of an SM singlet Dark Matter (DM) candidate in the Higgs sector. We will briefly comment on possible DM phenomenology in this work. For more extensive studies we refer, for example, to \cite{Bell:2016ekl,Arcadi:2022lpp}. The extension of the Higgs sector with a singlet has the advantage that it is possible to accommodate a light state, interpreted as the di-tau/di-photon resonance responsible for the experimental excesses under scrutiny, while having the rest of the BSM scalars at a heavier scale to comply with possible complementary bounds from other LHC searches and flavour physics. A similar result is much more difficult to achieve in the case of a Higgs sector composed of only two doublets as the constraints from requiring unitarity and boundedness from below of the scalar potential, as well as constraints from Electroweak Precision Tests (EWPTs), would disfavor very large mass splittings among the extra scalars besides the 125 GeV Higgs.
\\

\noindent This paper is structured as follows. In Sec. \ref{sec:pheno} we discuss the widths and production cross sections of a new (pseudo)scalar resonance using a model-independent approach, and discuss the couplings to SM particles the new (pseudo)scalar needs to have in order to reproduce the observed excesses. In Section \ref{sec:models} we describe the 2HDM+a and 2HDM+s models and their theoretical and experimental constraints. In Sec. \ref{sec:interp} we try to interpret the excesses in the specific setup given by these 2 models, and in Sec. \ref{sec:conclusion} we present our concluding remarks.

\section{Phenomenology of a 95 GeV (Pseudo)Scalar}
\label{sec:pheno}

\subsection{Minimal Flavour Violation and Natural Flavour Conservation}

Flavour violating processes, especially Flavour Changing Neutral Currents (FCNC), are strongly suppressed in the Standard Model. This is due to the fact that the large flavour group in the SM is broken only by the Yukawa couplings which, apart from the top-quark coupling, are relatively small. When extending the scalar sector of the Standard Model (SM), one needs to avoid generating large flavour changing processes mediated by neutral currents. This is done by enforcing that the pattern of the flavour violating structure in a BSM model remains the same as it is in the Standard Model. A necessary, but not sufficient, condition to obtain this is that the couplings of the new scalars are proportional to the Yukawa couplings of the SM. This condition is usually called Natural Flavour Conservation (NFC). This scenario is naturally realised when the couplings of the new scalars are acquired via mixing with the SM Higgs, for example in models where the additional scalar is a singlet \citep{Duerr:2016tmh,Khoze:2015sra,Baek:2015lna,Bauer:2016gys,Robens:2016xkb,Wang:2015cda,Costa:2015llh,Dupuis:2016fda,Balazs:2016tbi,Ko:2016ybp}.  
When extending the scalar sector with additional Higgs doublets it is possible, in general, to have a different Yukawa matrix for each doublet. To have, at tree level, the same Yukawa breaking structure as in the SM, one needs to assume that the Yukawa matrices of the new doublets are proportional to the SM one, with a different proportionality coefficient for each of them. This implementation of NFC is usually called ``The Aligned 2HDM" \citep{Pich:2009sp,Tuzon:2010vt}. However, no symmetry prevents deviation of the Yukawa matrix of the additional doublet from the SM structure via radiative corrections.
\\
\\
To avoid this problem one has to rely on the stronger hypothesis of Minimal Flavour Violation (MFV) \citep{D'Ambrosio:2002ex,Buras:2010mh}. This is concretely implemented by imposing a $\mathcal{Z}_2$ symmetry forbidding the two Higgs doublets to couple arbitrarily with all the SM fermions. There are four possible ways to impose such discrete symmetries, leading to the the Type-I, -II, -X and -Y 2HDMs \citep{Branco:2011iw}. These same configurations can also be defined in the 2HDM+s and 2HDM+a as the additional scalars are SM singlets.
\\
\\
In this section, we will start by making some general considerations regarding the required features of these new supposed scalars. We will only assume the NFC scenario, and we will set the Yukawa couplings of the new resonance to be proportional to the SM ones through some unconstrained proportionality constants $\xi_i$. In the rest of the paper we will then focus on the 2HDM+s and 2HDM+a models that assume a Yukawa sector with the MFV hypothesis, therefore working in the four $\mathcal{Z}_2$ symmetric Yukawa types.

\subsection{Fit of the signal} 
\label{2.2}
In this subsection we will illustrate, from a general perspective, our procedure to interpret the experimental excesses. We will refer to the following simplified lagrangians:
\begin{align}
    & \mathcal{L}=-\sum_f\frac{m_f}{v}\xi_f \bar f f \phi + 2g_{sVV}\frac{M_W^2W_\mu^+W^{-\mu}+\frac{1}{2}M_Z^2Z_\mu Z^\mu}{v}\phi\\
    & \mathcal{L}=-\sum_f\frac{m_f}{v}i \xi_f \bar f \gamma_5 f \phi 
\end{align}
describing, respectively, a scalar and pseudoscalar resonance (a sum over all SM quarks and charged leptons $f$ is assumed). As evident the Yukawa couplings are normalized, via scaling factors $\xi_f$, to the SM Yukawa couplings. In the scalar lagrangian we have also explicitly accounted for a coupling of the new scalar with the gauge bosons as it is generally present in the 2HDM+s. These couplings are taken to be  the ones of the Higgs but rescaled by a factor $g_{sVV}$. In the next section the scalar and pseudoscalar states will be identified, respectively, with the CP-even and CP-odd bosons belonging to the spectra of the 2HDM+s and 2HDM+a.

To interpret the $\gamma \gamma$ and $\tau \tau$ excesses with the models under consideration we will make use of the signal strength variables:
\begin{eqnarray}
    \mu_{\tau \tau} &=& \frac{\sigma_{\phi}\times BR(\phi\rightarrow \tau\tau)}{\left(\sigma_{\phi}\times BR(\phi\rightarrow \tau\tau)\right)_{SM}},\\
    \mu_{\gamma \gamma} &=& \frac{\sigma_{\phi}\times BR(\phi\rightarrow \gamma\gamma)}{\left(\sigma_{\phi}\times BR(\phi\rightarrow \gamma\gamma)\right)_{SM}}.
\end{eqnarray}
These are the values of the cross-sections times branching ratios of the decays in the considered final state channel of a resonance $\phi$, normalized by the counterpart for a SM-Higgs boson with the same mass as the considered resonance. We also define: 
\begin{eqnarray}
    R_{\tau\gamma} = \frac{\mu_{\tau\tau}}{\mu_{\gamma\gamma}} \rightarrow \frac{BR(\phi\rightarrow \tau\tau)}{BR(\phi\rightarrow \gamma\gamma)},
\end{eqnarray}
where the RHS holds when the same production mode is assumed in both cases. In this case, this parameter provides a particularly useful experimental constraint: it is unaffected by the specific production mode of the new scalar, only considering the ratio of the decay widths to ditau or diphoton final states. As such, the constraints from it can be applied to a model regardless of the relevant production modes.

The cross section for a resonant production process with final state X, where a spin-0 resonance $\phi=S (\mbox{scalar}),P (\mbox{pseudoscalar})$ of mass $M$ is produced and then decays, can be written as

\begin{eqnarray}
\sigma(p p \rightarrow \phi \rightarrow X) &=& \frac{\Gamma(\phi\rightarrow X)}{M \Gamma s} \sum_{i} C_{i} \Gamma(\phi\rightarrow i) = \frac{1}{M  s} \sum_{i} C_{i} \Gamma(\phi\rightarrow i) BR(\phi\rightarrow X),
\end{eqnarray}

\noindent where $i$ are the possible initial states, $C_i$ are weight factors that account for the proton PDFs and colour factors, and $s$ is the center of mass energy squared $s=(13TeV)^2$. The values of the $C_i$ are obtained from the PDFs as follows:

\begin{eqnarray}
C_{gg} &=& \frac{\pi^2}{8} \int_{M^2/s}^1 \frac{dx}{x} g(x)g\left(\frac{M^2}{sx}\right),\\
C_{q\bar{q}} &=& \frac{4\pi^2}{9} \int_{M^2/s}^1 \frac{dx}{x}\left(q(x)\bar{q}\left(\frac{M^2}{sx}\right) + q\left(\frac{M^2}{sx}\right)\bar{q}(x)\right).
\end{eqnarray}
\noindent 
For SM-like bosons, gluon fusion is typically the dominant production channel, i.e. $\sum_i C_i \simeq C_{gg}$. This may not be case in an extended Higgs sector as one can have enhanced couplings with the bottom quark, leading to a sizable contribution from $b$-fusion. We will assume, however, that this is not the case for the $\gamma\gamma$ excess as the CMS search for production in association with $b\bar{b}$ \citep{CMS:2023yay} does not show any anomaly with respect to the SM prediction. Assuming hence that the BSM resonance is produced mostly via gluon fusion and using the analytic expression above, the $\tau \tau$ signal strength can be rewritten as:
\begin{equation}
    \mu_{\tau \tau}=\frac{\sigma_{\phi}\times BR(\phi\rightarrow \tau\tau)}{\left(\sigma_{\phi}\times BR(\phi\rightarrow \tau\tau)\right)_{SM}}=R_{gg}R_{\tau \tau}
\end{equation}
with:
\begin{eqnarray}
    R_{gg} &=& \frac{\Gamma(\phi\rightarrow gg)}{\Gamma(\phi\rightarrow gg)_{SM}},\\
    R_{\tau\tau} &=& \frac{BR(\phi\rightarrow \tau\tau)}{BR(\phi\rightarrow \tau\tau)_{SM}}.
\end{eqnarray}
We adopt the value:
\begin{eqnarray}
    \left(\sigma_{gg\phi}\times BR(\phi\rightarrow \tau\tau)\right)_{SM} &=& 6.36pb,
\end{eqnarray}
based on the the MSTW2008 pdf set \citep{Martin:2009iq}, we get $\mu_{\tau \tau}=1.22$. Notice that the relations above hold for both a BSM scalar and pseudoscalar resonance.
\\
\\
For the case of the $\gamma \gamma$ signal we will also account for other production modes, and so the signal strength will be defined as:
\begin{eqnarray}
    \mu_{\gamma \gamma} &=& R_{gg} R_{\gamma\gamma} \frac{\sigma_{gg\phi,SM}}{\sigma_{\phi,SM}} \quad (P),\\
    \mu_{\gamma \gamma} &=& \frac{R_{gg}\sigma_{gg\phi, SM}+R_V\sigma_{VBF+VH,SM}+R_{tt}\sigma_{tt\phi,SM} }{\sigma_{\phi, SM}}R_{\gamma\gamma} \quad (S),
\end{eqnarray}
where $R_{\gamma\gamma}$ is the ratio of branching fraction compared to the SM, and $R_V, R_{tt}$ are, respectively, the ratios of the VBF+VH and $t\bar{t}$ associated production cross sections compared to the SM, 
\begin{eqnarray}
    R_{\gamma\gamma} &=& \frac{BR(\phi\rightarrow \gamma\gamma)}{BR(\phi\rightarrow \gamma\gamma)_{SM}},\\
    R_V &=& \frac{g^2_{VV\phi}}{g^2_{VVh}} = g^2_{sVV},\\
    R_{tt} &=& \frac{g^2_{tt\phi}}{g^2_{tth}} = \xi_u^2.
\end{eqnarray}
Notice that here the cases of a pseudoscalar and scalar resonance are distinct. In the former case, only the gluon fusion mode is available and hence we find a very similar expression to $\mu_{\tau \tau}$, but with a further scaling associated to the production cross-section of the resonance. For the scalar, we instead need to take the weighted average of the production modes, including the VBF and VH associated production modes. We also include the small contribution from production in association with a $t\bar{t}$ pair. This production mode is in principle also present for the pseudoscalar, but as it is further suppressed we choose to neglect it. Using the following values for the cross-sections of SM-like resonances \citep{LHCHiggsCrossSectionWorkingGroup:2016ypw}:
\begin{eqnarray}
    \sigma_{VBF+VH,SM}(13TeV) &=& 10.4pb,\\
    \sigma_{tt\phi,SM}(13TeV) &=&1pb,\\
    \sigma_{\phi,SM}(13TeV) &=& \sigma_{gg\phi,SM}+\sigma_{VBF+BH,SM}+\sigma_{tt\phi,SM} \sim 87.7pb,
\end{eqnarray}
the signal strength can be expressed in terms of the decay widths of the scalar and pseudoscalar states of the 2HDM+a/s. 
\\
\\
Finally, for the $b\bar{b}$ excess at LEP, the relevant observable is the ratio
\begin{eqnarray}
    \mu_{bb} &=& \frac{\sigma_{Z\phi}\times BR(\phi\rightarrow bb)}{\left(\sigma_{Z\phi}\times BR(\phi\rightarrow bb)\right)_{SM}} = R_{ZZ} R_{bb},
\end{eqnarray}
with 
\begin{eqnarray}
    R_{bb} &=& \frac{BR(\phi\rightarrow bb)}{BR(\phi\rightarrow bb)_{SM}},\\
    R_{ZZ} &=& \frac{\sigma_{Z\phi}}{\sigma_{Z\phi,SM}} = g_{\phi VV}^2.
\end{eqnarray}

\noindent The BSM decay widths entering in the signal strengths can be obtained in terms of the decay widths of SM-like resonances, via a suitable scaling procedure \citep{Denner:2011mq,Choi:2021nql}.
In the case of the 2HDM+s we have:
\begin{eqnarray}
    \Gamma(S\rightarrow bb) &=& \xi_{d}^2 \Gamma_{bb},\\
    \Gamma(S\rightarrow cc) &=& \xi_{u}^2 \Gamma_{cc},\\
    \Gamma(S\rightarrow \tau\tau) &=& \xi_{l}^2 \Gamma_{\tau\tau},\\
    \Gamma(S\rightarrow gg) &=& \frac{|\xi_{u} F_S\left(\frac{M^2}{4m_t^2}\right)+\xi_{d} F_S\left(\frac{M^2}{4m_b^2}\right)|^2}{| F_S\left(\frac{M^2}{4m_t^2}\right)+ F_S\left(\frac{M^2}{4m_b^2}\right)|^2} \Gamma_{gg},\\
    \Gamma(S\rightarrow \gamma\gamma) &=& \frac{|\frac{8}{3}\xi_{u} F_S\left(\frac{M^2}{4m_t^2}\right)+\frac{2}{3}\xi_{d} F_S\left(\frac{M^2}{4m_b^2}\right)+2\xi_{l} F_S\left(\frac{M^2}{4m_\tau^2}\right)-g_{sWW}F_W\left(\frac{M^2}{4m_W^2}\right)|^2}{| \frac{8}{3}F_S\left(\frac{M^2}{4m_t^2}\right)+ \frac{2}{3}F_S\left(\frac{M^2}{4m_b^2}\right)+2F_S\left(\frac{M^2}{4m_\tau^2}\right)-F_W\left(\frac{M^2}{4m_W^2}\right)|^2} \Gamma_{\gamma\gamma},
\end{eqnarray}
where the form factors $F_S, F_W$ are given in App. \ref{sec:decay}. For the SM-like decay widths $\Gamma_{ii} \equiv\Gamma(S\rightarrow ii)_{\rm SM}$ we have adopted the following values:
\begin{eqnarray}
    \Gamma_{bb} &=& 1.8847\MeV,\\
    \Gamma_{cc} &=& 0.0952\MeV,\\
    \Gamma_{\tau\tau} &=& 0.1955\MeV,\\
    \Gamma_{gg} &=& 0.1584\MeV,\\
    \Gamma_{\gamma\gamma} &=& 0.0032665\MeV,
\end{eqnarray}
with the $WW,ZZ,$ and $\gamma Z$ channels yielding a negligible contribution. Summing all the channels together one gets a total width $\Gamma=2.337\,\mbox{GeV}$.

In calculating the gluon and photon widths we have assumed that there is no other colored/charged particle that yields a sizeable contribution to the width. Of particular note is that in a model with an additional Higgs doublet one will necessarily have a contribution from the additional charged scalar. 

In this work, we have neglected the contributions from the diagram with the charged scalar loop, assuming that it must be small for two main reasons. First, the contribution from such a diagram is proportional to $g_{sWW}$, which must be small from Higgs physics constraints on VBF couplings. Second, the charged scalar must be very heavy due to flavour constraints as well as EWPT and unitarity constraints: as the additional neutral scalars need to be heavy, EWPT and unitarity will require a nearly-degenerate spectrum. Moreover, if the charged scalar would yield a sizeable contribution to the $\gamma\gamma$ width of the new scalar, it would do the same for the diphoton partial width of the $125\GeV$ Higgs boson, which is also constrained. While current constraints still provide some freedom for small additional contributions, in view of the other constraints from VBF, flavour physics, and unitarity, we have decided not to consider such a scenario. 
\\
\\
Following an analogous line of reasoning, the decay width of a pseudoscalar resonance, belonging to the mass spectrum of the 2HDM+a, can be written as \citep{Choi:2021nql}: 

\begin{eqnarray}
    \Gamma(P\rightarrow bb) &=& \xi_{d}^2 \frac{\Gamma_{bb}}{1-\frac{4m_b^2}{M^2}} \frac{1+\delta_P^{bb}}{1+\delta_S^{bb}} \sim 0.99 \times \xi_{d}^2 \Gamma_{bb}, \\
    \Gamma(P\rightarrow cc) &=& \xi_{u}^2 \frac{\Gamma_{cc}}{1-\frac{4m_c^2}{M^2}} \frac{1+\delta_P^{cc}}{1+\delta_S^{cc}} \sim 1.03 \times \xi_{u}^2 \Gamma_{cc}, \\
    \Gamma(P\rightarrow \tau\tau) &=& \xi_{l}^2 \frac{\Gamma_{\tau\tau}}{1-\frac{4m_\tau^2}{M^2}} \sim  \xi_{l}^2 \Gamma_{\tau\tau}, \\
    \Gamma(P\rightarrow gg) &=& \frac{|\xi_{u} F_P\left(\frac{M^2}{4m_t^2}\right)+\xi_{d} F_P\left(\frac{M^2}{4m_b^2}\right)|^2}{| F_S\left(\frac{M^2}{4m_t^2}\right)+ F_S\left(\frac{M^2}{4m_b^2}\right)|^2} \frac{1+\delta_P^{gg}}{1+\delta_S^{gg}}\Gamma_{gg},\\
    \Gamma(P\rightarrow \gamma\gamma) &=& \frac{|\frac{8}{3}\xi_{u} F_P\left(\frac{M^2}{4m_t^2}\right)+\frac{2}{3}\xi_{d} F_P\left(\frac{M^2}{4m_b^2}\right)+2\xi_{l} F_P\left(\frac{M^2}{4m_\tau^2}\right)|^2}{| \frac{8}{3}F_S\left(\frac{M^2}{4m_t^2}\right)+ \frac{2}{3}F_S\left(\frac{M^2}{4m_b^2}\right)+2F_S\left(\frac{M^2}{4m_\tau^2}\right)-F_W\left(\frac{M^2}{4m_W^2}\right)|^2} \nonumber\\
    &\times& \frac{1+\delta_P^{\gamma\gamma}}{1+\delta_S^{\gamma\gamma}}\Gamma_{\gamma\gamma},
\end{eqnarray}
where the form factor $F_P$ is given in App. \ref{sec:decay} and the correction factors $\delta_{P,S}^{ff,gg,\gamma\gamma}$ are from \citep{Choi:2021nql}. The pseudoscalar has no tree-level couplings to the $WW$ and $ZZ$ channels. Similarly, the pseudoscalar does not couple at tree-level to the charged scalar; 
hence no contribution to the effective coupling with $\gamma \gamma$ is induced.

Combining the previous expressions we can finally write the $R_{ii}$ factors for a scalar resonance:
\begin{eqnarray}
    R_{gg} &=& \frac{|\xi_{u} F_S\left(\frac{M^2}{4m_t^2}\right)+\xi_{d} F_S\left(\frac{M^2}{4m_b^2}\right)|^2}{| F_S\left(\frac{M^2}{4m_t^2}\right)+ F_S\left(\frac{M^2}{4m_b^2}\right)|^2},\\
    R_{\gamma\gamma} &=& \frac{|\frac{8}{3}\xi_{u} F_S\left(\frac{M^2}{4m_t^2}\right)+\frac{2}{3}\xi_{d} F_S\left(\frac{M^2}{4m_b^2}\right)+2\xi_{l} F_S\left(\frac{M^2}{4m_\tau^2}\right)-g_{sWW}F_W\left(\frac{M^2}{4m_f^2}\right)|^2}{| \frac{8}{3}F_S\left(\frac{M^2}{4m_t^2}\right)+ \frac{2}{3}F_S\left(\frac{M^2}{4m_b^2}\right)+2F_S\left(\frac{M^2}{4m_\tau^2}\right)-F_W\left(\frac{M^2}{4m_f^2}\right)|^2},
\end{eqnarray}
and a pseudoscalar resonance:
\begin{eqnarray}
    R_{gg} &=& \frac{|\xi_{u} F_P\left(\frac{M^2}{4m_t^2}\right)+\xi_{d} F_P\left(\frac{M^2}{4m_b^2}\right)|^2}{| F_S\left(\frac{M^2}{4m_t^2}\right)+ F_S\left(\frac{M^2}{4m_b^2}\right)|^2} \frac{1+\delta_P^{gg}}{1+\delta_S^{gg}},\\
    R_{\gamma\gamma} &=& \frac{|\frac{8}{3}\xi_{u} F_P\left(\frac{M^2}{4m_t^2}\right)+\frac{2}{3}\xi_{d} F_P\left(\frac{M^2}{4m_b^2}\right)+2\xi_{l} F_P\left(\frac{M^2}{4m_\tau^2}\right)|^2}{| \frac{8}{3}F_S\left(\frac{M^2}{4m_t^2}\right)+ \frac{2}{3}F_S\left(\frac{M^2}{4m_b^2}\right)+2F_S\left(\frac{M^2}{4m_\tau^2}\right)-F_W\left(\frac{M^2}{4m_f^2}\right)|^2} \frac{1+\delta_P^{\gamma\gamma}}{1+\delta_S^{\gamma\gamma}}.
\end{eqnarray}

It is important to remark that the above procedure is valid under the implicit assumption that the scalar or pseudoscalar resonance has no extra exotic contributions to its width, such as, for example, decays into DM pairs.
To justify this assumption we have conducted a study by performing a numerical fit of the anomalies, considering the scaling factors $\xi_f$ as well as a hypothetical additional branching fraction $BR_{\rm inv}$ as free parameters.

\begin{figure}
    \centering
    \includegraphics[width = 0.49\textwidth]{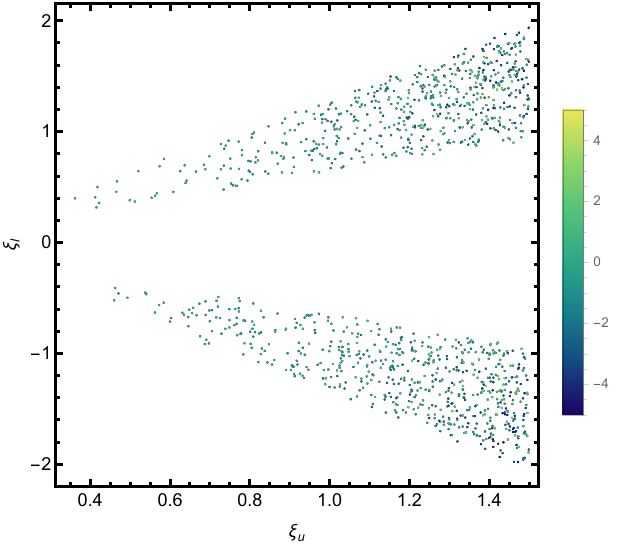}
    \includegraphics[width = 0.49\textwidth]{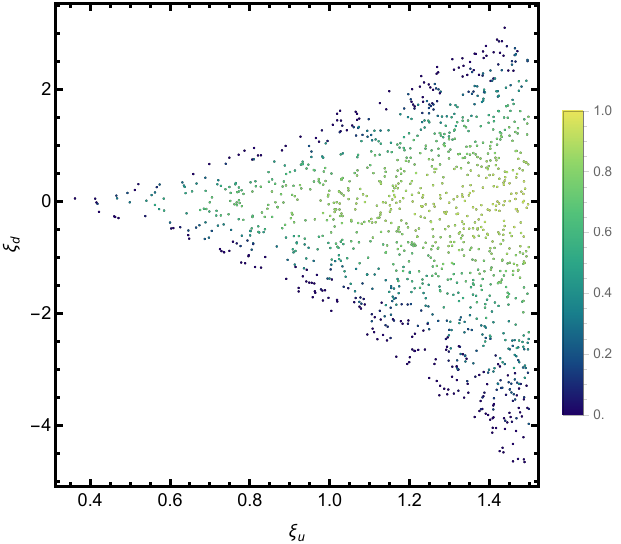}
    \caption{Point scan showing where a pseudoscalar resonance with $\xi_i$ coupling multipliers is compatible with CMS+ATLAS excesses at $2\sigma$ C.L.. Left: $\xi_u$ vs $\xi_l$, the color code indicates the value of $\xi_b$. Right: $\xi_u$ vs $\xi_d$, the color code indicates the value of the invisible BR.
    }
    \label{fig:genPS}
\end{figure}

Fig. \ref{fig:genPS} shows the case of a pseudoscalar resonance. Here the 4-dimensional parameter space $(\xi_u,\xi_d,\xi_l,\textit{BR}_{inv})$ has been considered. We have performed a scan on the $\xi_f$'s over the following ranges: 
\begin{eqnarray}
    0<&\xi_u&<1.5,\\
    -10<&\xi_d&<10,\\
    -10<&\xi_l&<10,
\end{eqnarray}

and considered, for each assignation, the optimal value of the invisible branching fraction to reproduce both experimental excesses. The points that reproduce both the $\tau\tau$ and $\gamma\gamma$ signals at $2\sigma$ C.L. are shown in the plot.
In the case of the pseudoscalar, thanks to the larger couplings that arise at loop-level to gluons and photons, it is usually not a problem to have a large enough production rate to fit the signals. It is therefore useful to assume a non-zero invisible branching ratio. In the left panel of Fig. \ref{fig:genPS} we see how $\xi_u$ and $\xi_l$, which are the two main couplings that one needs to chose to reproduce the signal, need to be related to reproduce the right signal strengths in both channels. 

In the right panel, we can instead see how the coupling to bottom quarks is correlated to the invisible branching ratio: to avoid signal strengths that are too large one needs to have a branching ratio to $\tau\tau$ similar to the standard model, and therefore there must be another partial decay width that dominates the total width, either to $b\bar{b}$ pairs or invisible products. 
\\

\begin{figure}
    \centering
    \includegraphics[width = 0.49\textwidth]{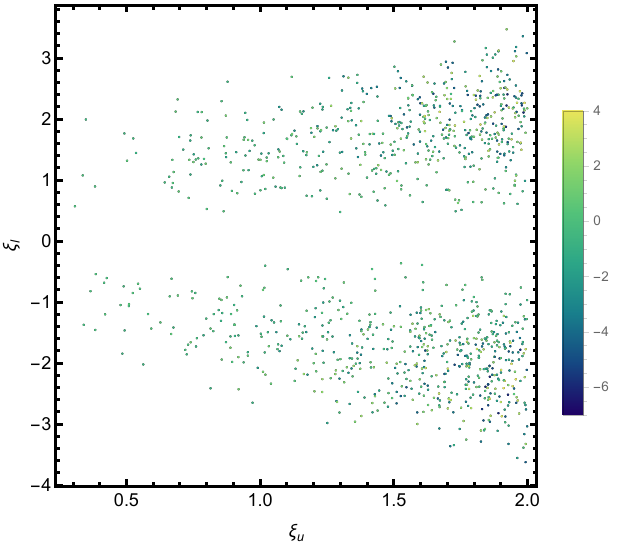}
    \includegraphics[width = 0.49\textwidth]{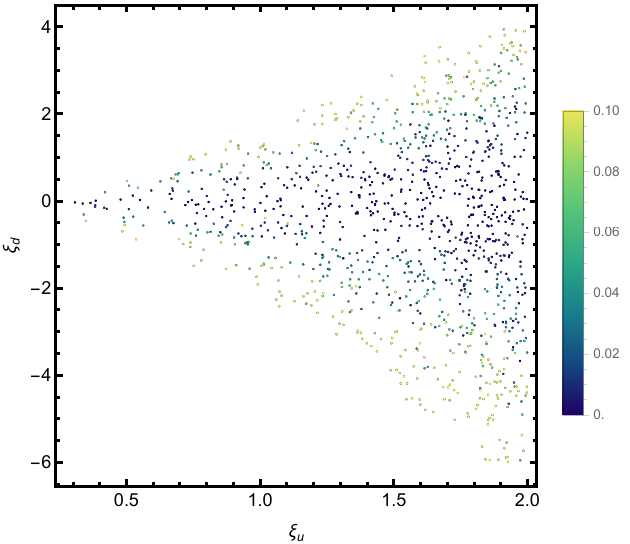}\\
    \includegraphics[width = 0.49\textwidth]{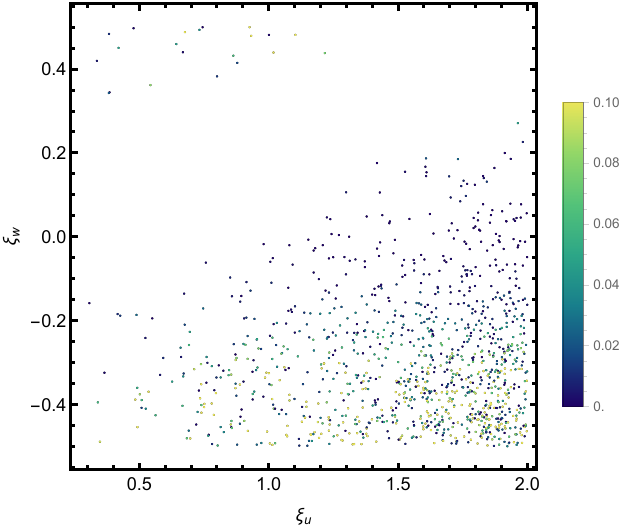}
    \caption{Point scan showing where a Scalar resonance with $\xi_i$ coupling multipliers is compatible with CMS+ATLAS excesses at $2\sigma$ C.L.. Top Left: $\xi_u$ vs $\xi_l$, the color code indicates the value of $\xi_b$. Top Right: $\xi_u$ vs $\xi_d$, the color code indicates the value of $R_{bb}$. Bottom: $\xi_u$ vs $\xi_w$, the color code indicates the value of $R_{bb}$. 
    }
    \label{fig:genS}
\end{figure}

\noindent 
In Fig. \ref{fig:genS} we show analogous results for a scalar resonance. Here the parameter space is 5-dimensional as we have added a further scaling parameter $\xi_w$ to account for coupling of the resonance to the $W$ boson. Again we have performed a scan over the coupling modifiers: 
\begin{eqnarray}
    0<&\xi_u&<2,\\
    -10<&\xi_d&<10,\\
    -10<&\xi_l&<10,\\
    -0.5<&\xi_w&<0.5,
\end{eqnarray}
and adapted \textit{a posteriori} $BR_{inv}$ to properly fit the experimental signal. We do not require the points to also reproduce the LEP signal. Unlike in the pseudoscalar case, it is usually difficult to have a production rate that is too large to fit the signal, and therefore most of the points in our scan have invisible branching fractions smaller than $1\%$. In the top left panel of Fig. \ref{fig:genS} we see how $\xi_u$ and $\xi_l$, which are, again, the two main couplings that one needs to chose to reproduce the signal, need to be related to reproduce the right signal strengths in both channels. The result is similar to the case of a pseudoscalar. 

In the top right panel, we can instead see how the coupling to bottom quarks is correlated to the signal strength for $b\bar{b}$ at LEP. One usually needs couplings at the 2 extremes of the interval to get the right signal at LEP. This is due to the fact that for the values of $\xi_w$ chosen to be in agreement with Higgs physics constraints, the production rate is barely sufficient to get a signal strength at the lower end of the signal strength interval with a branching fraction close to $100\%$. This will be a general requirement, in the case of the scalar, if we want to be able to also reproduce the signal at LEP within the allowed regions of parameter space. 

Finally, in the bottom panel we see the result of the scan in the plane $\xi_u$ vs $\xi_w$. We can see that, in general, negative values\footnote{Actually, the requirement is to have opposite sign compared to $\xi_u$.} of the $W$ coupling are preferred to generate constructive interference in the photon loop. The branching fraction to photons, for the scalar case, is in fact usually too small to reproduce $R_{\gamma \gamma}$.

\section{The 2HDM+s and the 2HDM+a models}
\label{sec:models}

\subsection{2HDM+a: particle content and scalar potential}
To describe the 2HDM+a we adopt the following potential for the scalar sector of the theory\cite{Abe:2018bpo}:
\begin{equation} 
V_{\rm 2HDM\!+\!a} = V_{\rm 2HDM}\! + \! \frac{1}{2} m_{a^0}^2 (a^0)^2\!+\! \frac{\lambda_a}{4} (a^0)^4
\!+\! \left(i \kappa a^0 \Phi^{\dagger}_1\Phi_2\!+\!\mbox{h.c.}\right)
+ \left(\lambda_{1P}(a^0)^2 \Phi_1^{\dagger}\Phi_1 \!+\! \lambda_{2P}(a^0)^2 \Phi_2^{\dagger}\Phi_2\right), 
\label{eq:V2HDa}
\end{equation} 
where $\Phi_{1,2}$ are the two Higgs doublets and $a^0$ is a pseudoscalar SM singlet state. $V_{\text{2HDM}}$ is the conventional (softly-broken) $Z_2$ symmetric 2HDM potential:
\begin{align}
 V_{\rm 2HDM} &= m_{11}^2 \Phi_1^\dagger \Phi_1+ m_{22}^2 \Phi_2^\dagger \Phi_2 - m_{12}^2  (\Phi_1^\dagger \Phi_2 + {\rm h.c.} ) +\frac12{\lambda_1} ( \Phi_1^\dagger \Phi_1 )^2 +\frac12{\lambda_2} ( \Phi_2^\dagger \Phi_2 )^2  \nonumber \, \\ &
+\lambda_3 (\Phi_1^\dagger \Phi_1) (\Phi_2^\dagger \Phi_2) +\lambda_4 (\Phi_1^\dagger \Phi_2 ) (\Phi_2^\dagger \Phi_1) +\frac12 {\lambda_5} [\,   (\Phi_1^\dagger \Phi_2 )^2 + {\rm h.c.} \, ] \, .
\label{eq:scalar_potential}
\end{align}
After EW symmetry breaking the doublet fields can be rewritten as:
\begin{equation}
    \Phi_{i}=\left(
    \begin{array}{c}
         \phi_i^+  \\
          (v_i+\rho_i+i \eta_i)/\sqrt{2}
    \end{array}\right)
    \label{eq:doublets}
\end{equation}
with $\rho_i$ and $\eta_i$ being real scalar and pseudoscalar degrees of freedom respectively, while $\sqrt{v_1^2+v_2^2}=v\simeq 246\,\mbox{GeV}$. From them it is possible to define physical degrees of freedom:
\begin{align}
& \left(
    \begin{array}{c}
         \phi_1^+  \\
         \phi_2^+ 
    \end{array}
    \right)=\mathcal{R}_{-\beta} \left(
    \begin{array}{c}
         G^+  \\
         H^+ 
    \end{array}
    \right),\,\,\,
    \left(
    \begin{array}{c}
         \rho_1 \\
         \rho_2 
    \end{array}
    \right)=\mathcal{R}_{-\alpha-\frac{\pi}{2}} \left(
    \begin{array}{c}
         h_{125}  \\
         H 
    \end{array}
    \right),\nonumber\\
    & \left(
    \begin{array}{c}
         \eta_1  \\
         \eta_2 
    \end{array}
    \right)=\mathcal{R}_{-\beta} \left(
    \begin{array}{c}
         G^0  \\
         A^0 
    \end{array}
    \right),\,\,\,\,\left(
    \begin{array}{c}
         A^0  \\
         a^0 
    \end{array}
    \right)=\mathcal{R}_{\theta} \left(
    \begin{array}{c}
         A  \\
         a 
    \end{array}
    \right),\,\,\,\,
\end{align}
where $\mathcal{R}_\psi$ is the general $2\times 2$ rotation matrix: 
\begin{equation}
    \mathcal{R}_\psi=\left(
    \begin{array}{cc}
      \cos\psi   &  \sin\psi  \\
       -\sin\psi  &  \cos\psi 
    \end{array}
    \right).
\end{equation}
The angles $\alpha,\beta$ are the mixing angles conventionally defined in two-Higgs doublet models, with $\tan\beta=v_2/v_1$. The angle $\theta$ is an additional mixing angle for the pseudoscalar states defined by (we customarily assume the $m_a < m_A$ hierarchy):
\begin{equation}
\sin2\theta=\frac{2 \kappa v}{m_{A}^2-m_{a}^2}\;.\label{eq:sin2theta}
\end{equation}
\\

The full parameter space of the model is thus comprised of 13 parameters
\begin{equation}
    m_{11}^2, m_{12}^2, m_{22}^2, m_{a^0}^2, k, \lambda_1, \lambda_2, \lambda_3, \lambda_4, \lambda_5, \lambda_{1P}, \lambda_{2P}, \lambda_a\label{eq:set1}
\end{equation} 
that can be rewritten in terms of
\begin{equation}
    v, m_h, m_H, m_A, m_a,\allowbreak m_{H^\pm}, \tan\beta, \cab, \theta, \lambda_3, \lambda_{P1}, \lambda_{P2}\label{eq:set2}
\end{equation}

\noindent Of such 13 parameters, the Higgs vev and $h_{125}$ mass ($v,m_h$) are fixed by the known values, leaving 11 degrees of freedom. One can also optionally add a DM candidate as in \cite{Abe:2018bpo} with the addition of 2 more parameters. 
The 2HDM+a parameter space is therefore high-dimensional. Moreover, it is difficult to uniformly sample it, especially in our limits of interest (the alignment and/or decoupling limits), because in such case the relations between different sets of parameters like the ones in Eq.\ref{eq:set1} and Eq. \ref{eq:set2}  can involve fine-tuned relations. Our approach is therefore to identify the parameter regions that are most likely to satisfy all constraints and where one can predict signal strengths analytically to compare them with the experimental results.
\\
We will see that it is necessary for the other new scalars $H,A,H^{\pm}$ to be heavy ($\gtrsim \mathcal{O}(\TeV)$). In this case, the $\gamma \gamma, \tau \tau, b \bar{b}$ signals arising from the production and decay of the light resonance $a$, along with all other relevant constraints involving the production of $a$ alone, is sensitive to a small subset of the parameter space ($m_a, \tan\beta, \cab, \theta$).

We will fix the value of $m_a=95\GeV$, assuming that, taking into account detector resolution, all experimental excess peaks are compatible with this value, as done in \citep{Belanger:2012tt,Becirevic:2015fmu,Cao:2016uwt,Fox:2017uwr,Haisch:2017gql,Heinemeyer:2021msz,Azevedo:2023zkg,Biekotter:2023jld,Biekotter:2023oen,Dutta:2023cig,Aguilar-Saavedra:2023vpd,Borah:2023, Iguro:2022dok,Iguro:2022fel,Ashanujjaman:2023etj,Cao95GeV,Cao:2019ofo,Bhattacharya:2023lmu}. In our study of the 2HDM+a we also assume the model satisfies the alignment limit, enforcing this by setting $\beta - \alpha = \pi/2$. This ensures that the mass eigenstate $h_{125}$, identified with the 125 GeV Higgs, features SM-Higgs-like couplings, thus automatically satisfying all Higgs physics constraints related to $h_{125}$ only. The assumption of the alignment limit has a negligible impact on the interpretation of the experimental excesses as the $\tau \tau$ and $\gamma \gamma$ signal strengths depend, at least at LO, only on $\tan \beta$ and $\sin\theta$. This will be not the case in the 2HDM+s, where will consider $\beta-\alpha$ as free in this latter case, as clarified in the following section.

Thus, our relevant parameter space for the 2HDM+a interpretation is comprised of just 2 degrees of freedom, the parameters $\tan\beta$ and $\theta$. From Eq. \ref{eq:sin2theta} we see that $\theta$ can be defined up to a shift of $\pi$, for example between $-\pi/2$ and $\pi/2$, with the sign defined by the relative sign of the coupling $\kappa$ and the mass hierarchy $m_A^2-m_a^2$. 
Without loss of generality we will adopt the range $\left[0,\frac{\pi}{4}\right]$ for the variation of the angle $\theta$. In such a way the hierarchy $m_a < m_A$ is always maintained with the lightest state being mostly singlet-like.

\subsection{2HDM+s: particle content and scalar potential}

The model with an additional scalar can be described with a scalar potential very similar to the case of the pseudoscalar:

\begin{equation} 
V_{\rm 2HDM\!+\!s} = V_{\rm 2HDM}\! + \! \frac{1}{2} m_{s^0}^2 (s^0)^2\!+\! \frac{\lambda_s}{4} (s^0)^4
\!+\! \left(\kappa s^0 \Phi^{\dagger}_1\Phi_2\!+\!\mbox{h.c.}\right)
+ \left(\lambda_{1S}(s^0)^2 \Phi_1^{\dagger}\Phi_1 \!+\! \lambda_{2S}(s^0)^2 \Phi_2^{\dagger}\Phi_2\right).
\label{eq:V2HDs}
\end{equation} 

\noindent After EW symmetry breaking the two doublets can be written as in Eq. \ref{eq:doublets}, while the singlet also gains a vev:

\begin{eqnarray}
    s_0 &=& v_s+s.
\end{eqnarray}
The charged scalar and pseudoscalar mass matrices can be diagonalised as in the 2HDM:

\begin{align}
& \left(
    \begin{array}{c}
         \phi_1^+  \\
         \phi_2^+ 
    \end{array}
    \right)=\mathcal{R}_{-\beta} \left(
    \begin{array}{c}
         G^+  \\
         H^+ 
    \end{array}
    \right),\,\,\,
    \left(
    \begin{array}{c}
         \eta_1 \\
         \eta_2 
    \end{array}
    \right)=\mathcal{R}_{-\beta} \left(
    \begin{array}{c}
         G^0  \\
         A 
    \end{array}
    \right).
\end{align}
However, the 2HDM+s has 3 scalars that in general will mix in a non-trivial way. It is convenient to rewrite the fields $\rho_i$ in terms of the states $h^0, H^0$:

\begin{align}
    & \left(
    \begin{array}{c}
         \rho_1  \\
         \rho_2 
    \end{array}
    \right)=\mathcal{R}_{-\beta} \left(
    \begin{array}{c}
         h^0  \\
         H^0 
    \end{array}
    \right).\,\,\,\,
\end{align}
These are the linear combinations that one would obtain by rotating the doublets to the Higgs basis.

Defining the rotation matrices

\begin{eqnarray}
    \mathcal{R}_\psi^{12}=\left(
    \begin{array}{ccc}
      \cos\psi   &  \sin\psi  & 0 \\
       -\sin\psi  &  \cos\psi & 0 \\
       0 & 0 & 1
    \end{array}
    \right),\quad \mathcal{R}_\psi^{13}=\left(
    \begin{array}{ccc}
      \cos\psi  & 0 &  -\sin\psi \\
      0 & 1 & 0 \\
       \sin\psi & 0 &  \cos\psi 
    \end{array}
    \right), \quad \mathcal{R}_\psi^{23}=\left(
    \begin{array}{ccc}
    1 & 0 & 0\\
      0 &\cos\psi  &   \sin\psi   \\
      0 & -\sin\psi &   \cos\psi 
    \end{array}
    \right),
\end{eqnarray}

\noindent the general rotation matrix that will diagonalise the mass matrix can be expressed in terms of 3 angles as

\begin{eqnarray}
    \mathcal{R}_\theta^{23} \mathcal{R}_{\delta}^{13} \mathcal{R}_\gamma^{12}, 
\end{eqnarray}

\noindent obtaining

\begin{align}
    & \left(
    \begin{array}{c}
         h^0  \\
         H^0 \\
         s^0
    \end{array}
    \right)=\mathcal{R}_\theta^{23} \mathcal{R}_{\delta}^{13} \mathcal{R}_\gamma^{12}  \left(
    \begin{array}{c}
         h_{125}  \\
         H\\
         s
    \end{array}
    \right).\,\,\,\,\label{eq:rotationscalar}
\end{align}

\noindent The 2HDM scenario is recovered for $\theta=\delta=0, \gamma= \beta-\alpha-\frac{\pi}{2}$, with the 2HDM alignment limit scenario recovered for all angles equal to zero. As we have seen, the benefit of interpreting the signals in terms of a scalar rather than a pseudoscalar is that the new scalar is also able to couple to the gauge bosons and reproduce the $\bar{b}b$ excess. To do this, however, we need to work strictly away from the alignment limit. 
\\
\\

\noindent As discussed in Section 2.2, it is usually difficult in the 2HDM+s to have sufficiently large production rates to reproduce the LEP $b\bar{b}$ signal. As such, we focus on a benchmark case that should provide a best-case scenario to reproduce this signal: 
\begin{eqnarray}
    \delta &=& -\left(\frac{\pi}{2}+\alpha-\beta\right),\\
    \gamma &=&0,
\end{eqnarray}
where the naming convention for $\delta$ is chosen as, for $\gamma=0$, the angle $-(\delta+\frac{\pi}{2})$ \ has the same impact as $\alpha-\beta$ has in the standard 2HDM. The expressions for the fields $h^0,H^0,s^0$ in this limit are:
\begin{eqnarray}
    h^0 &=& -\sab h_{125} +\cab s\label{eq:hstates},\\
    H^0 &=& -\cab \sin\theta h_{125} + \cos\theta H -\sab\sin\theta s\label{eq:scalarHcouplings},\\
    s^0 &=& -\cab \cos\theta h_{125} -\sin\theta H -\sab\cos\theta s.
\end{eqnarray}
Finally, the modifiers to the gauge boson couplings can be read from Eq. \ref{eq:hstates} as
\begin{eqnarray}
    g_{hVV} &=& -\sab,\\
    g_{sVV} &=& \cab.\label{eq:svvcoupling}
\end{eqnarray}

Like in the pseudoscalar case, the full parameter space of the model is comprised of 13 parameters $m_{11}^2, m_{12}^2,\allowbreak m_{22}^2, m_{s^0}^2, k, \lambda_1, \lambda_2, \lambda_3, \lambda_4, \lambda_5, \lambda_{1S}, \lambda_{2S}, \lambda_s$, with two specific combinations being fixed by the SM vev and $h_{125}$ mass, leaving 11 degrees of freedom. 
 
However, for the purpose of our analysis we care about processes where only the new light scalar $s$ is produced, and assume the other new scalars are heavy $\gtrsim \mathcal{O}(\TeV)$. In this case, as in the pseudoscalar case, the only relevant parameters for the analysis will be $m_s, \tan\beta, cos(\alpha-\beta),\theta$.

Likewise to the pseudoscalar case, we set $m_s=95\GeV$. As seen in Sec. \ref{2.2}, in the scalar case it can become challenging to obtain signals that are large enough to reproduce the observed rates. We therefore chose to restrict our analysis where the $\theta$ angle lies close to maximal mixing. Consequently we will adopt $\theta=\pm \pi/4$. Therefore our analysis will consider a parameter space comprised by just two degrees of freedom $\tan\beta, \cab$.

\subsection{Yukawa sectors}

The Yukawa lagrangian of the 2HDM+a for the neutral scalars,  in terms of the physical states, is:
\begin{equation}
\mathcal{L}_{\rm Yuk}=-\frac{m_f}{v}\sum_f \bigg[ g_{hff} h_{125} \bar f f+g_{Hff}
H\bar f f- i g_{Aff} A \bar f \gamma_5 f-i g_{aff} a \bar f \gamma_5 f  \bigg] \, , 
\end{equation}
where: 
\begin{align}
    & g_{Aff}=\cos\theta g_{A^0ff},\,\,\,\,g_{aff}=\sin\theta \, g_{A^0ff},\,\,\,
\end{align}
and the coefficients $g_{hff},g_{Hff},g_{A^0ff}$ are given in Tab. \ref{table:2hdm_cplgs}. In the alignment limit, with $\alpha\rightarrow\beta-\frac{\pi}{2}$, the modifiers $\xi_f^{H,A}$ depend only on $\tan\beta$, reducing to $g_{H,Aff}$: $\xi_f^{H}=g_{Hff},\xi_f^{A}=g_{A^0ff}$. 
\\
\\
For the 2HDM+s the Yukawa terms for the neutral scalars are:
\begin{equation}
\mathcal{L}_{\rm Yuk}=-\frac{m_f}{v}\sum_f \bigg[ g_{hff} h_{125} \bar f f+g_{Hff}
H\bar f f +g_{sff}s\bar f f - i g_{A^0ff} A \bar f \gamma_5 f  \bigg] \, , 
\end{equation}
where\footnote{\noindent Note that due to our sign convention, to get a similar interference pattern as in a pure 2HDM, one needs to choose a negative value of $\sin\theta$. Reversing the value of $\sin\theta$ is equivalent to flipping the sign of $\cab$, so we will only consider negative values of $\sin\theta$.}: 
\begin{eqnarray}
    g_{hff} &=&-\sab+\cab\sin\theta\xi_f^H,\\
    g_{Hff} &=&-\cos\theta \xi_f^H,\\
    g_{sff} &=&\cab + \sab\sin\theta \xi_f^H.\label{eq:sffcoupling}
\end{eqnarray}

\begin{table}
\renewcommand{\arraystretch}{1.17}
\begin{center}
\begin{tabular}{|c|c|c|c|c|c|}
\hline
~~~~~~ & ~~~~~~ &  Type I & Type II & Type X & Type Y \\ \hline \hline 
- & $g_{htt}$ & $ \frac{\cos \alpha} { \sin \beta} \rightarrow 1$ & $\frac{ \cos \alpha} {\sin \beta} \rightarrow 1$ & $\frac{ \cos \alpha} {\sin\beta} \rightarrow 1$ & $ \frac{ \cos \alpha}{ \sin\beta} \rightarrow 1$ \\ \hline
- & $g_{hbb}$ & $\frac{\cos \alpha} {\sin \beta} \rightarrow 1$ & $-\frac{ \sin \alpha} {\cos \beta} \rightarrow 1$ & $\frac{\cos \alpha}{ \sin \beta} \rightarrow 1$ & $-\frac{ \sin \alpha}{ \cos \beta} \rightarrow 1$ \\ \hline
- & $g_{h\tau\tau} $ & $\frac{\cos \alpha} {\sin \beta} \rightarrow 1$ & $-\frac{\sin \alpha} {\cos \beta} \rightarrow 1$ & $- \frac{ \sin \alpha} {\cos \beta} \rightarrow 1$ & $\frac{ \cos \alpha} {\sin \beta} \rightarrow 1$  \\ \hline\hline
$\xi_{u}$ & $g_{Htt}$ & $\frac{\sin \alpha} {\sin \beta} \rightarrow -\frac{1}{\tan\beta}$ & $\frac{ \sin \alpha} {\sin \beta} \rightarrow -\frac{1}{\tan\beta}$ & $ \frac{\sin \alpha}{\sin \beta} \rightarrow -\frac{1}{\tan\beta}$ & $\frac{ \sin \alpha}{ \sin \beta} \rightarrow -\frac{1}{\tan\beta}$ \\ \hline
$\xi_{d}$ & $g_{Hbb}$ & $ \frac{ \sin \alpha}{\sin \beta} \rightarrow -\frac{1}{\tan\beta}$ & $\frac{\cos \alpha}{\cos \beta} \rightarrow {\tan\beta}$ & $\frac{\sin \alpha} {\sin \beta} \rightarrow -\frac{1}{\tan\beta}$ & $\frac{ \cos \alpha} {\cos \beta} \rightarrow {\tan\beta}$ \\ \hline
$\xi_{l}$ & $g_{H\tau\tau}$ & $\frac{ \sin \alpha} {\sin \beta} \rightarrow -\frac{1}{\tan\beta}$ & $\frac{\cos \alpha} {\cos \beta} \rightarrow {\tan\beta}$ & $\frac{ \cos \alpha} {\cos \beta} \rightarrow {\tan\beta}$ & $\frac{\sin \alpha} {\sin \beta} \rightarrow -\frac{1}{\tan\beta}$ \\ \hline\hline
$\xi_{u}$ & $g_{A^0tt}$ & $\frac{1}{\tan\beta}$ & $\frac{1}{\tan\beta}$ & $\frac{1}{\tan\beta}$ & $\frac{1}{\tan\beta}$ \\ \hline
$\xi_{d}$ & $g_{A^0bb}$ & $-\frac{1}{\tan\beta}$ & ${\tan\beta}$ & $-\frac{1}{\tan\beta}$ & ${\tan\beta}$ \\ \hline
$\xi_{l}$ & $g_{A^0\tau\tau}$ & $-\frac{1}{\tan\beta}$ & ${\tan\beta}$ & ${\tan\beta}$ & $-\frac{1}{\tan\beta}$
\\ \hline
\end{tabular}
\vspace*{2mm}
\caption{Couplings of the 2HDM Higgs bosons to third generation fermions, normalized to the SM-Higgs couplings, as a function of the angles $\alpha$ and $\beta$ for the four types of 2HDM scenarios. For the CP-even $h,H$ states,  the values in the alignment limit $\alpha\! \to\! \beta\!-\!\frac{\pi}{2}$ are also shown.
}
\label{table:2hdm_cplgs}
\end{center}
\vspace*{-6mm}
\end{table}

\subsection{Theoretical constraints}


The couplings of the scalar potential are subject to a series of constraints from perturbativity, unitarity, and boundedness from below of the potential. The case of the 2HDM+a has been considered, for example, in \cite{Abe:2018bpo,Arcadi:2022lpp}. We list the resulting constraints below: 
\begin{itemize}
    \item {\it Unitarity:}
\begin{align}
\label{eq:unitarity}
    & |x_i| < 8\pi \, , \ |\lambda_{1,2P}|< 4\pi,\,\,\,\,|\lambda_3\pm \lambda_4|< 4 \pi
    \, , \nonumber\\ & 
    \left \vert \frac{1}{2}\left(\lambda_1+\lambda_2 \pm \sqrt{(\lambda_1-\lambda_2)^2+4\lambda_k^2}\right)\right \vert < 8 \pi \, , \  k=4,5 \, , \nonumber\\
    & |\lambda_3+ 2 \lambda_4 \pm 3 \lambda_5|< 8 \pi,\,\,\,\,|\lambda_3 \pm \lambda_5| < 8 \pi \, , 
\end{align}
where the $x_i$'s are the eigenvalues of the $2\rightarrow 2$ scattering matrices, and are given by the solutions of the equation 
\begin{align}
& 0=x^3-3 (\lambda_a+\lambda_1+\lambda_2)x^2  + (9 \lambda_1 \lambda_a+9 \lambda_2 \lambda_a -4 \lambda_{1P}^2-4 \lambda_{2P}^2-4 \lambda_3^2-4 \lambda_3 \lambda_4-\lambda_4^2+9 \lambda_1 \lambda_2)x \nonumber\\
& +12 \lambda_{2P}^2 \lambda_1+12 \lambda_{1P}^2\lambda_2 
 -16 \lambda_{1P}\lambda_{2P}\lambda_3-8 \lambda_{1P}\lambda_{2P}\lambda_4 +(-27 \lambda_1 \lambda_2+12 \lambda_3^2+12 \lambda_3 \lambda_4+3 \lambda_4^2)\lambda_a \, . 
\end{align}

\item {\it Boundedness from below (BFB):}

\begin{equation}
\label{eq:boundedbelow}
\begin{split}
  & \lambda_{1} > 0, \quad \lambda_{2} > 0, \quad \lambda_{a} > 0,   \\
  & \bar{\lambda}_{12} \equiv \sqrt{\lambda_{1} \lambda_{2}}+\lambda_{3}+\min (0, \lambda_{4}- |\lambda_{5}|) > 0,   \\
  & \bar{\lambda}_{1P} \equiv \sqrt{\frac{\lambda_{1} \lambda_{a}}{2}} + \lambda_{1P} > 0,   \ \ \  
  \bar{\lambda}_{2P} \equiv \sqrt{\frac{\lambda_{2} \lambda_{a}}{2}} + \lambda_{2P} > 0,   \\
  & \sqrt{\frac{\lambda_{1} \lambda_{2} \lambda_{a}}{2}}   + \lambda_{1P} \sqrt{\lambda_{2}} + \lambda_{2P} \sqrt{\lambda_{1}} + [\lambda_{3} + \min (0, \lambda_{4} - |\lambda_{5}|)] \sqrt{\frac{\lambda_{a}}{2}}
  + \sqrt{2} \sqrt{\bar{\lambda}_{12} \bar{\lambda}_{1P} \bar{\lambda}_{2P}} > 0.
  \end{split}
\end{equation}

\end{itemize}

Furthermore one has to ensure that the global minimum of the scalar potential coincides with the EW vacuum and, in particular, no vacuum expectation value is acquired by the field $a_0$. For a more detailed discussion we refer, for example, to \cite{Abe:2019wjw}.
\\
\\
The constraints for the scalar model, with the potential of Eq. \ref{eq:V2HDs}, can be obtained directly by making the substitutions $\lambda_{iP}\rightarrow\lambda_{iS}$ and $\lambda_P\rightarrow\lambda_S$ in the above expressions.

\subsection{Experimental constraints}

\subsubsection{Flavour constraints}

We consider 2HDM Yukawa sectors protected by a $\mathcal{Z}_2$ symmetry that ensures the absence of tree level off-diagonal couplings of neutral gauge bosons with quarks. This symmetry ensures that such couplings are not generated even by the RGE running of the Yukawa couplings at lower energies. However, as the theory includes an additional charged scalar, Flavour Changing Neutral Currents (FCNCs) appear necessarily at loop level. The overall good agreement of measured flavour observables with SM predictions requires therefore that the contributions from new physics are small. The most relevant flavour constraints for a 2HDM are \citep{Enomoto:2015wbn}: the leptonic meson decays $M\rightarrow l\nu$, $B_q^0\rightarrow \mu^+\mu^-$, the hadronic $\tau$ lepton decays $\tau\rightarrow M\nu$, the radiative $B$ meson decay $B\rightarrow X_s \gamma$, the mass splitting due to neutral $B_q^0$ meson mixing  $\Delta M_q$, and the $\epsilon_K$ parameter measuring $CP$ violation in the $K^0-\bar{K}^0$ system.
\\
\\
The leptonic decays  $M\rightarrow l\nu$ and the $\tau$ hadronic decays $\tau\rightarrow M\nu$ happen at tree level. The BSM physics contribution is on the order of 
\begin{eqnarray}
    \mathcal{O}\left(\left(\frac{m_u\xi_u\xi_l-m_d\xi_d\xi_l }{m_u+m_d}\right)\frac{m_M^2}{m_{H^+}^2}\right) \approx \mathcal{O}\left(\xi_d\xi_l \frac{m_M^2}{m_{H^+}^2}\right).
\end{eqnarray}
These contributions are especially relevant for the Type II 2HDM at large $\tan\beta$. In our case, as we will consider a very heavy charged scalar and focus on low $\tan\beta$, we expect these decays to be subdominant with respect to the other constraints.\\

Regarding the leptonic decays $B_q^0\rightarrow \mu^+\mu^-$, the most constraining one is $B_s^0\rightarrow \mu^+\mu^-$. Recently, new updated experimental values of this observable gave results closer to the Standard Model expectation value \citep{LHCb:2021vsc,LHCb:2021awg,CMS:2022mgd}. We use the current world average from \citep{HeavyFlavorAveragingGroup:2022wzx}. To get a theoretical prediction we use the fitting formula from \citep{Enomoto:2015wbn} for the Standard Model result, calculated at NNLO QCD \citep{Hermann:2013kca} and NLO EW \citep{Bobeth:2013tba,Bobeth:2013uxa}. We include the one-loop BSM physics contribution adapting the result for the $\mathcal{Z}_2$ symmetric 2HDM from \citep{Cheng:2015yfu,Enomoto:2015wbn}.\\

The inclusive radiative $B$ meson decay $B\rightarrow X_s \gamma$ ($b\rightarrow s\gamma$) can suffer from large QCD and non-perturbative corrections. For this reason, we use the Standard Model prediction from \citep{Misiak:2015xwa,Czakon:2015exa, Enomoto:2015wbn}, which includes QCD corrections at NNLO and calculable long-distance effects. For the BSM physics contribution, we use the expression quoted in \citep{Enomoto:2015wbn} and based on \citep{Kagan:1998ym,Hurth:2003dk,Lunghi:2006hc} at NNLO in QCD. For the experimental value, we use the world average from \citep{HeavyFlavorAveragingGroup:2022wzx}.\\

The neutral $B_q^0$ meson mixings $\Delta M_s$ and $\Delta M_d$ are measured very precisely \citep{HeavyFlavorAveragingGroup:2022wzx}, but suffer from larger theoretical uncertainties because they require input parameters calculated on lattice \citep{Dowdall:2019bea,Bazavov:2017lyh}. We use the expressions of the Wilson coefficients from \citep{Enomoto:2015wbn} and use the 4 loop top $\overline{MS}$ mass from \citep{Marquard:2015qpa}. 
For the $\epsilon_K$ $CP$-violating parameter we start from the traditional approach of \citep{Enomoto:2015wbn,Brod:2010mj,Brod:2011ty}, which expresses the full amplitude in terms of $tt$, $tc$, and $cc$ terms by incorporating the result of the $u$ diagrams using the CKM matrix identity $\sum_i V_{is}V_{id}^\star=0$. We then update to the approach of \cite{Brod:2019rzc,Brod:2022har}, where instead the same identity is used to recast the sum in terms of $tt$, $tu$, and $uu$ terms. This latter approach has the advantage of a better-converging, more well-behaved perturbative expansion. We use the bag parameter from \citep{FlavourLatticeAveragingGroup:2019iem}, and include long-distance corrections not included in the bag parameter with the additional factor $k_e$ from \citep{Buras:2010pza}. The experimental result is taken from \citep{Enomoto:2015wbn}. All remaining particle data is taken from \citep{ParticleDataGroup:2022pth}.
\\
\\
When deriving flavour constraints for BSM physics, one needs to be careful about the values of the CKM matrix elements used. The standard values quoted are obtained in fits that usually include the values of the observables $\Delta M_{s,d},\epsilon_K$, and assume Standard Model relations between the CKM elements and the observables. When deriving constraints on BSM physics that affects flavour observables, one therefore needs to either make a new fit including all observables using the relations between parameters and observables given by the BSM model, or fit the CKM matrix elements by only using observables that are not affected by BSM physics and then use the obtained CKM elements to impose constrains on the observables affected by BSM physics. This is particularly important for $\epsilon_K$, whose expression also relies on the CKM orthogonality relation. In this work, we have chosen the latter approach. 
 
We use CKMfitter \citep{Charles:2004jd} to obtain the Wolfenstein parameters without including the $\Delta M_{s,d},\epsilon_K$ observables in the fit. We use the latest data from \citep{ParticleDataGroup:2022pth,HeavyFlavorAveragingGroup:2022wzx} for the CKM matrix elements $|V_{ud}|$, $|V_{cb}|$, $|V_{ub}|$, $|V_{ub}|f_+$, the branching ratios $\mathcal{B}(\tau\rightarrow K\nu)$, $\mathcal{B}(K\rightarrow l\nu)$, the unitarity triangle angles $\cos2\beta$, $\sin2\beta$, $\alpha$, $\gamma$, and $\alpha_S(M_Z)$. To have the unitarity and orthogonality relations hold precisely, we recast the Wolfenstein parameters into $\sin\theta_{12}$, $\sin\theta_{23}$, $\sin\theta_{13}$, and $\delta$. The results are shown in Tab. \ref{table:ckmparam}.
\\
\\

\begin{table}
\renewcommand{\arraystretch}{1.17}
\begin{center}
\begin{tabular}{|c|c|c|c|}
\hline\hline
$\lambda$ & $A$ &  $\rho$ & $\eta$ \\  \hline 
$0.225$ & $0.8234$ & $0.143$ & $0.348$ \\ \hline\hline
$\sin\theta_{12}$ & $\sin\theta_{23}$ &  $\sin\theta_{13}$ & $\delta$ \\  \hline 
$0.225$ & $0.04168$ & $0.003529$ & $1.181$ \\ \hline\hline 
\end{tabular}
\vspace*{2mm}
\caption{Paramaters of CKM matrix used for flavour constraints.}
\label{table:ckmparam}
\end{center}
\vspace*{-6mm}
\end{table}

\noindent Defining:
\begin{eqnarray}
\tilde{\mathcal{R}}_{\psi,\delta}^{13}=\left(
    \begin{array}{ccc}
      \cos\psi  & 0 &  \sin\psi e^{-i \delta} \\
      0 & 1 & 0 \\
       -\sin\psi e^{i\delta} & 0 &  \cos\psi 
    \end{array}
    \right),
\end{eqnarray}

\noindent The resulting CKM matrix we use is:

\begin{eqnarray}
    V_{CKM} &=& \mathcal{R}_{\theta_{23}}^{23} \tilde{\mathcal{R}}_{\theta_{13},\delta}^{13}\mathcal{R}_{\theta_{12}}^{12}\\ 
    &=& \left(
    \begin{array}{ccc}
      0.9744  & 0.2250 &  0.001341 - 0.003264 i \\
      -0.2249 - 0.0001326 i & 0.9735 - 0.00003061 i & 0.04168 \\
       0.008073 - 0.003177 i & -0.04092 - 0.0007337 i &  0.9991 
    \end{array}
    \right).
\end{eqnarray}


\subsubsection{EWPT constraints}

The presence of additional scalars breaks the custodial symmetry of the SM Higgs sector. Such violation can be expressed in terms of a deviation $\Delta \rho$ of the $\rho$ parameter. In the case of the 2HDM+a:
\begin{align}
\Delta \rho_{\text{2HDM+a}} & = \frac{\alpha_{\rm QED} (m_Z^2)}{16 \pi^2 m_W^2 (1 -m_W^2/m_Z^2)}\bigg \{ \big[ 
f(m^2_{H\pm},m^2_H) + \cos^2\theta f(m^2_{H\pm},m^2_A) \nonumber \\ 
&  + \sin^2\theta f(m^2_{H\pm},m^2_a) - \cos^2\theta f(m^2_A,m^2_H) - \sin^2\theta f(m^2_a,m^2_H) \big]  \bigg \},
\end{align}
where $\alpha_{\rm QED}$ is the fine structure constant evaluated at $m_Z$ and the loop function $f$ is:
\begin{equation}
f(x,y) = x+y- \frac{2 x y}{x-y} \log \frac{x}{y} \, . 
\end{equation}
As $\lim_{y\rightarrow x}f(x,y)=0$, it can be easily seen that if $m_H=m_A=m_{H^{\pm}}$ then $\Delta \rho=0$, i.e. the SM result is recovered. If this is not the case, very strong constraints apply to the model.
\\
\\
For the scalar model, considering our rotation matrix of Eq. \ref{eq:rotationscalar}, we have:
\begin{align}
\Delta \rho_{\text{2HDM+s}} & = \frac{\alpha_{\rm QED} (m_Z^2)}{16 \pi^2 m_W^2 (1 -m_W^2/m_Z^2)}\bigg \{ \big[ 
f(m^2_{H\pm},m^2_A) + \cos^2\theta f(m^2_{H\pm},m^2_H)  \nonumber \\ 
& + \sin^2\theta\sab^2 f(m^2_{H\pm},m^2_s) + \sin^2\theta\cab^2 f(m^2_{H\pm},m^2_h) - \cos^2\theta f(m^2_A,m^2_H) \nonumber\\ 
& - \sin^2\theta\sab^2 f(m^2_A,m^2_s)- \sin^2\theta\cab^2 f(m^2_A,m^2_h) \nonumber\\
&+3\cab^2\left(f(m_Z^2,m_s^2)-f(m_Z^2,m_h^2)-f(m_W^2,m_s^2)+f(m_W^2,m_h^2)\right)\big]  \bigg \}.
\end{align}

In this case, in the degenerate limit $m_H=m_A=m_{H^{\pm}}$ only the terms in the last row survive, and one gets
\begin{eqnarray}
    \Delta\rho &\sim& 0.011 \cab^2 \alpha_{\rm QED} (m_Z^2),
\end{eqnarray}
which will be much smaller than the SM contribution. We also notice that $\Delta \rho=0$ in the alignment limit.
\\
\\
To account for EWPT constraints, we have performed, along the same lines of \cite{Arcadi:2022lpp}, a $\chi^2$ analysis of the Peskin $S$, $T$, and $U$ parameters \cite{Kanemura:2011sj,Haber:2010bw}:
\begin{equation}
\label{eq:chi2}
\chi^2=\sum_{i,j}( {\cal O}_i-{\cal O}_i^{\rm SM}){\left(\sigma_i V_{ij} \sigma_j\right)}^{-1}( {\cal O}_j-{\cal O}_j^{\rm SM}) \, , 
\end{equation}
where ${\cal O}=(S,T,U)$ are given by
\begin{eqnarray}
    S &=& -\frac{1}{4\pi}\left(g(m^2_{H\pm},m^2_{H\pm})-\cos^2\theta g(m^2_A,m^2_H)- \sin^2\theta g(m^2_a,m^2_H)\right),\\
    T &=& \frac{\Delta\rho}{\alpha_{\rm QED}},\\
    U &=& -\frac{1}{4\pi}\left(g(m^2_{H\pm},m^2_{H\pm})+\cos^2\theta g(m^2_H,m^2_A)+ \sin^2\theta g(m^2_H,m^2_a)\right.\nonumber\\
    &&\left.-g(m^2_{H\pm},m^2_H)-\cos^2\theta g(m^2_{H\pm},m^2_A)- \sin^2\theta g(m^2_{H\pm},m^2_a)\right),
\end{eqnarray}
for the pseudoscalar model, while the expressions for the scalar model can be derived using \citep{Haber:2010bw}. 

The function $g$ is given by
\begin{eqnarray}
    g(x,y) &=& -\frac{1}{3}\left(\frac{4}{3}-\frac{x\log x - y \log y}{x-y}-\frac{x+y}{(x-y)^2}f(x,y)\right).
\end{eqnarray}

${\cal O}^{\rm SM}$ represent the central values in the case of the SM \cite{Baak:2011ze,Baak:2014ora,Haller:2018nnx}:
\begin{equation}
 {\cal O}^{\rm SM}= (S,T,U)^{\rm SM} =  
 (0.04,0.09,-0.02)\, .
\end{equation}
For the standard deviations and covariance matrices we have taken the following values:
\begin{equation}
\sigma=(0.11,~0.14,~0.11) \ , \ \ \   V= \left(
\begin{array}{ccc} 1 & 0.92 & -0.68 \\ 0.92 & 1 & -0.87 \\ -0.68 & -0.87 & 1 \end{array} \right) \, .
\label{eq:covariance}
\end{equation}

\subsubsection{Higgs physics constraints}

Since we are considering extended Higgs sectors, we must consider a series of constraints from the measurements of the properties of the 125 GeV Higgs at the LHC. We might first consider the possibility of additional 2-body decays, for example $h_{125} \rightarrow \phi \phi, \phi=a,s$. These are, however, not relevant for our study as $m_\phi=95\,\mbox{GeV}$ and such 2-body decays are kinematically forbidden. One might, however, have 3-body decays, $h_{125}\rightarrow \phi \phi^* \rightarrow \phi \bar f f$ \footnote{In presence of DM one might also consider $h_{125} \rightarrow \bar \chi \chi \phi$. As already pointed out we have neglected this case of study.}.

In the 2HDM+a in the alignment limit the decay rate $h_{125}\rightarrow a \bar f f$ is written as:
\begin{eqnarray}
\Gamma(h_{125} \rightarrow a \bar f f)=\frac{N_c^f |g_{A^0ff}|^2}{128\pi^3}\frac{m_f^2}{v^2}\frac{|g_{haa}|^2}{m_h} g\left(\frac{4 m_a^2}{m_h^2}\right) \sin^2 \theta,
\end{eqnarray}

where:
\begin{eqnarray}
g(x)=\frac{1}{8}(x-4)\left[4-\log\left(\frac{x}{4}\right)\right]-\frac{5x-4}{4 \sqrt{x-1}}\left[\arctan\left(\frac{x-2}{2\sqrt{x-1}}\right)-\arctan\left(\frac{1}{\sqrt{x-1}}\right)\right].
\end{eqnarray}
We can compare this with the direct decay $h_{125}\rightarrow \bar{f}f$:
\begin{eqnarray}
    \frac{\Gamma(h_{125} \rightarrow a \bar f f)}{\Gamma(h_{125} \rightarrow\bar f f)} &=& \frac{ 1}{16\pi^2}\frac{|g_{haa}|^2}{m_h^2} g\left(\frac{4 m_a^2}{m_h^2}\right)\sin^2 \theta.
\end{eqnarray}
The coupling $g_{haa}$ has a non-trivial dependence on the parameters of the potential. We provide its complete expression in App. \ref{sec:decay}. Its value is constrained by unitarity limits, and taking it to be $\mathcal{O}(m_h)$, we get
\begin{eqnarray}
    \frac{\Gamma(h_{125} \rightarrow a \bar f f)}{\Gamma(h_{125} \rightarrow\bar f f)} &\sim& 10^{-7} \left(\frac{g_{haa}}{m_h}\right)^2.
\end{eqnarray}
We therefore do not expect this decay to impose  any relevant constraint. A similar result is found for the 2HDM+s. For this reason we do not show this case explicitly. 

The other relevant constraints come from the signal strengths of the $h_{125}$. As will be clarified below, the alignment limit will be strictly adopted for the 2HDM+a; consequently the latter constraints are automatically overcome. In the case of the 2HDM+s we have adapted the constraints on the so called universal coupling-strength scale factors $\kappa_F$ (associated to coupling to SM fermions) for all fermions and $\kappa_V$ (associated to couplings to vector bosons) determined, for example, in \citep{ATLAS:2015egz,ATLAS:2015ciy}. Notice that this procedure is valid as, in our setup, the branching ratios of decay processes of the $h_{125}$ for non-SM final states are negligible.

We implement the 2HDM+s model in \HiSi \citep{Bahl:2022igd,Bechtle:2020uwn}, using the latest available version 1.1.3. We check that the $2\sigma$ C.L. regions obtained with \HiSi are compatible with the ones we obtain by the reinterpretation of \citep{ATLAS:2015ciy}. We also check that our exclusion regions and the relevant searches we consider match the bounds and exclusion regions obtained by \HiBo \citep{Bahl:2022igd,Bechtle:2020pkv}. This includes the collider bounds listed in the next section.

\subsubsection{Collider constraints}

The 2HDM+a and 2HDM+s have rich collider phenomenology, with the main searches that apply to these models being resonance searches and MET searches. The main missing energy searches are Monojet $j+\met$ \citep{CMS:2023xlp,ATLAS:2021kxv}, Mono-$\gamma+\met$ \citep{CMS:2023xlp,ATLAS:2020uiq}, Mono-$W+\met$ \citep{CMS:2017zts,ATLAS:2018nda}, Mono $Z+\met$ \citep{ATLAS:2018nda,CMS:2020ulv}, Mono-Higgs $h_{125}+\met$ \citep{CMS:2019ykj,ATLAS:2023ild,ATLAS:2021shl} and $t\bar{t}/b\bar{b}+\met$ \citep{ATLAS:2022ygn}. For our type of models, the Mono-$Z$, Mono-Higgs, and $t\bar{t}+\met$ are typically the most constraining, due to better sensitivity in these channels. All these searches, however, apply only if the new scalar has invisible decay channels. While for a pseudoscalar it is easy to interpret the signal while also allowing for a sizeable invisible branching fraction, we have seen in Sec. \ref{sec:pheno} in the model with a scalar it becomes challenging. Therefore, in our 2HDM+a/s interpretations we will strictly assume that the new particle has no invisible decays. This can happen if the dark matter particle usually considered in these models is above the kinematic threshold for decay ($m_\chi>m_\phi/2$), or if the coupling to the dark matter particle is vanishing small. 
\\
\\
We consider all relevant resonance searches for the lightest additional scalar, all of which have an excess. The supposed particle is below the $t\bar{t}$ threshold. The only search considering a $\bar{b}b$ final state in this mass region is the one at LEP with its $\sim2\sigma$ excess. An excess is also seen in the $\tau\tau$ and $\gamma\gamma$ final states. Dijet resonance searches are usually performed in much higher mass ranges due to the large QCD background at low invariant masses. Finally, searches for production of the scalar in association with $t\bar{t}/b\bar{b}$ \citep{CMS:2022arx} are relevant for the lightest (pseudo)scalar, and will be included in our analysis. 
\\
\\
Additional resonance searches are possible for the heavier scalars through cascade decays. If a DM particle is present and below the kinematic threshold for the heavier scalars to decay into it, these processes will also generate $\met$ signatures. These resonance cascade searches look for processes such as $H\rightarrow aZ,h_{125}h_{125},h_{125}s,ss,aa$, \citep{CMS:2019ogx,ATLAS:2022hwc,ATLAS:2021ifb,CMS:2022qww}, $A\rightarrow ah_{125},Zh_{125},Zs$ \citep{CMS:2019qcx}, and $H^\pm\rightarrow W^\pm s, W^\pm h_{125}, W^\pm a, tb$ \citep{CMS:2022jqc,ATLAS:2021upq}. These searches will, in general, set a lower bound on the mass of the heavy scalars at about $1\TeV$. Resonance cascade searches have good sensitivity if the mass gaps in the particle spectrum allow for on-shell production of the heavy scalars, and if the heavier states are not so heavy such that their production rates are very small.  
\\
\\
The ATLAS search $H^\pm \rightarrow tb$ \citep{ATLAS:2021upq} is interpreted in the context of the 2HDM, and sets a lower bound on the mass of the charged scalar $m_{H^\pm}\gtrsim 800\GeV$ at $\tan\beta=1$ in the 2HDM in the alignment limit. In the 2HDM+a and 2HDM+s there are additional decay channels for $H^\pm$ and, depending on the branching fractions, the 2HDM+a/s could feature a less-constraining lower bound on the mass of the charged scalar. The CMS search $H^\pm\rightarrow\phi W^\pm$ \cite{CMS:2022jqc} does not apply to our scenario, as the mass of the neutral boson is set to $200\GeV$, and the search does not find any excess for $m_{H^\pm}\le700\GeV$ anyway. As flavour constraints force the mass of the charged scalar to lie above $1\TeV$, these searches do not add additional constraints to our analysis.
\\
\\
In the 2HDM+a model with alignment, in the region of parameter space of interest for this work, $A$ decays predominantly in the $t\bar{t}$ channel, while $H$ decays predominantly in either the $aZ$ or $t\bar{t}$ channel. The  $H \rightarrow h_{125}h_{125}$ decay channel vanishes in the alignment limit, and the $aa$ decay channel vanishes for $\tan\beta=1$ and $\lambda_{1P}=\lambda_{2P}$ and is in any case subleading in the decoupling regime $m_H\gg m_a$. Therefore, the most constraining search for our 2HDM+a interpretation is the CMS search $H\rightarrow aZ$ \citep{CMS:2019ogx}. This search, interpreted in the context of a 2HDM, excludes $m_H\lesssim650\GeV$ for $m_a\sim 95\GeV$   and $\tan\beta=1.5$. As the reinterpretation of these searches in the 2HDM+a/s usually leads to less stringent constraints, this search does not add additional constraints to our analysis.
\\
\\
Another search applicable to the 2HDM+a in the alignment limit, but which has low sensitivity in our region of parameter space, is $H\rightarrow aa\rightarrow b\bar{b}b\bar{b}$ \citep{CMS:2022qww}. Given the low expected $BR(H\rightarrow aa)$, this search does not have the necessary sensitivity to exclude any mass hypothesis in the considered range for $m_X=m_H$   between $1\TeV$ and $3\TeV$ in our analysis setup. 
\\
\\
The most relevant search for the 2HDM+s is the $A\rightarrow Zs$ process \citep{CMS:2019ogx}, which requires $m_A\gtrsim 700\GeV$ for $m_s\sim95\GeV$, $\tan\beta=1.5$, and $\cab=0.01$. Additionally, there are searches that are relevant only away from alignment. The search $H\rightarrow h_{125}h_{125}$ \citep{ATLAS:2022hwc} explores the mass range $250\GeV<m_H<5\TeV$. Assuming $\sigma_{Hgg}=(\sigma_{Hgg})_{SM}$ and $BR(H\rightarrow h_{125}h_{125})= 0.05$, this search excludes scalars in the mass range $350\GeV<m_H<1\TeV$ at 2 sigma. Assuming instead $BR(H\rightarrow h_{125}h_{125})= 0.1$, the mass range $320\GeV<m_H<1.3\TeV$ is excluded. The search $H\rightarrow ss\rightarrow b\bar{b}b\bar{b}$ \citep{CMS:2022qww} considers the mass windows $1\TeV<m_H<3\TeV$, excluding scalars in the mass range $1.1\TeV<m_H<1.35\TeV$ at 2 sigma for $\sigma_{Hgg}=(\sigma_{Hgg})_{SM}$, $BR(s\rightarrow b\bar{b})= BR(s\rightarrow b\bar{b})_{SM}$, and $BR(H\rightarrow ss)= 0.05$. For $BR(H\rightarrow ss)= 0.1$, the mass range $1.1\TeV<m_H<1.45\TeV$ is instead excluded. The search $A\rightarrow Zh_{125}$ \citep{CMS:2019qcx} excludes $m_A\lesssim 2m_t$. Interestingly, the searches \citep{CMS:2022suh,ATLAS:2022hwc,CMS:2022qww}, which consider 4-$b$ final states, have a similar $\sim2\sigma$ excess in correspondence of $m_H=1.6\TeV.$ 
\\
\\
Overall, we can expect to be able to evade all collider constraints with the exception of $t\bar{t}\phi$ \citep{CMS:2022arx} by assuming that the heavy scalars have a mass $m_{H,A,H^\pm}\gtrsim 1\TeV$ in the 2HDM+a or a mass $m_{H,A,H^\pm}\gtrsim 1.4\TeV$ in the 2HDM+s. These assumptions are similar to those needed to satisfy flavour constraints. The search $t\bar{t}\phi$, where the light (pseudo)scalar is produced in association with a $t\bar{t}$ pair, is potentially sensitive enough to exclude some parts of the parameter space we are interested in. This search targets $\tau^+\tau^-$ final states for the decay of the (pseudo)scalar $\phi$, so the sensitivity will mostly depend on the value of the branching fraction and therefore on the chosen Yukawa type.

Note Added: a few months after the publication of the first version of this manuscript, a new analysis from CMS of $t\bar{t}\phi,\phi\rightarrow\tau\tau$ and $W\phi,\phi\rightarrow\tau\tau$ events was published \citep{CMS:2024ulc}. The limits included in this analysis are a significant improvement comparing to \citep{CMS:2022arx} and, as we will see, change the conclusions of our analysis significantly, at least for the scalar case.

\section{Interpreting the excesses}
\label{sec:interp}

\subsection{Interpreting the excesses in the 2HDM+a model}
\label{sec:2hdmaint}

\begin{figure}
    \centering
    \includegraphics[width = 0.49\textwidth]{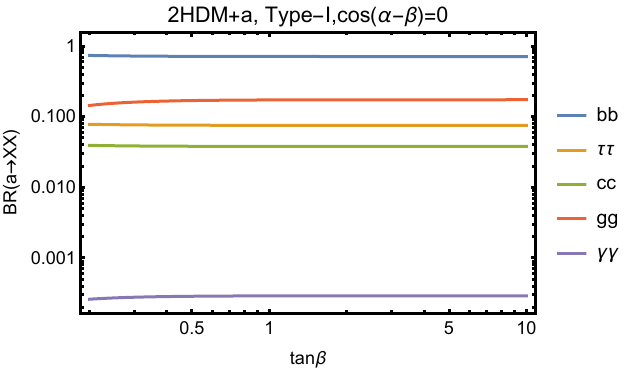}
    \includegraphics[width = 0.49\textwidth]{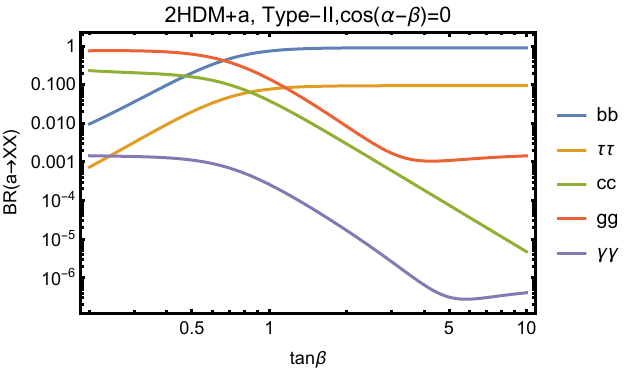}\\
    \includegraphics[width = 0.49\textwidth]{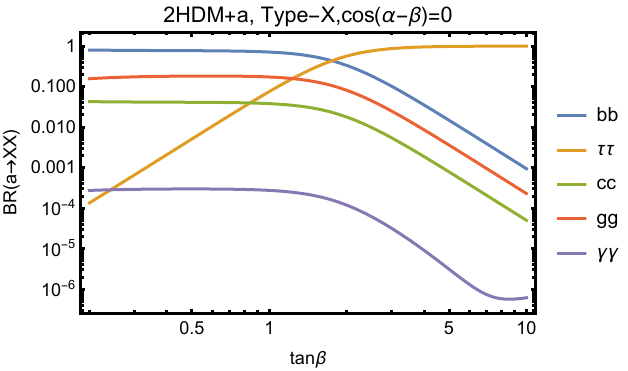}
    \includegraphics[width = 0.49\textwidth]{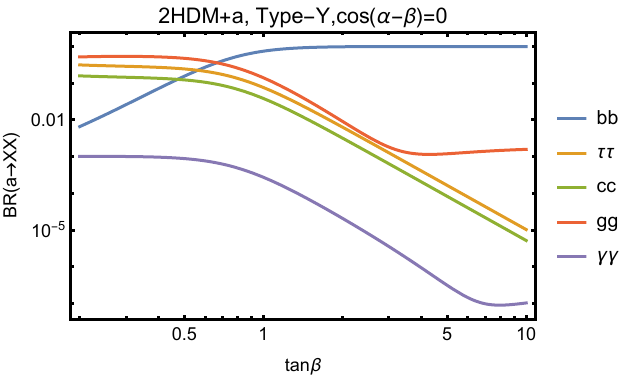}
    \caption{Decay branching ratios of the pseudoscalar $a$ particle for the different coupling configurations: Type I (top left), Type II (top right), Type X (bottom left), Type Y (bottom right). Invisible BR is assumed to be zero, and we assume the alignment limit $\cab=0$. 
    }
    \label{fig:adecay}
\end{figure}

To interpret the observed excesses in terms of the 2HDM+a, we start by computing the branching ratios of the $95\GeV$ pseudoscalar. These are shown in Fig.  \ref{fig:adecay} for the four types of 2HDM Yukawa couplings. We assume no invisible decay width.

Next, we use the signal strengths described in Sec. \ref{2.2} to identify the regions of parameter space of interest. We start from $R_{\tau\gamma}$. As all decay widths to SM particles are proportional to $\sin^2\theta$, the $\theta$ dependence cancels out in $R_{\tau\gamma}$ in the 2HDM+a. Neglecting the effect of $b$ and $\tau$ in the loop diagrams that couple $a$ to $\gamma\gamma$, the value of $R_{\tau\gamma}$ is 

\begin{eqnarray}
    R_{\tau\gamma} &\sim& \frac{4.6\xi_l^2}{\xi_u^2} = \left\{ \begin{array}{ c l }
    4.6 & \quad \textrm{Type I and Y, }\\
    4.6\tan^4\beta   & \quad \textrm{Type II and X. }
  \end{array}\right.
\end{eqnarray}

\noindent For the pseudoscalar case this must satisfy
\begin{eqnarray}
    2 \lesssim \frac{\sigma_{h,SM}}{\sigma_{ggh,SM}}R_{\tau\gamma}\lesssim 10.8,
\end{eqnarray}

\noindent which means

\begin{eqnarray}
    1.7 \lesssim R_{\tau\gamma}\lesssim 9.4.
\end{eqnarray}

\noindent We identify the following $\tan\beta$ regions where this variable allows compatibility of the model with the observed excesses: 
\begin{itemize}
    \item For type I and Y, any value of $\tan\beta$ is compatible with the ratio of the signals,
    \item For type II and X, one needs $0.7\lesssim\tan\beta\lesssim1.2$.
\end{itemize}
Moreover, the central value for $R_{\tau\gamma}$ is
\begin{eqnarray}
    R_{\tau\gamma}\sim 3.94,
\end{eqnarray}
which is very close to the value it has for Type I and Y, and can be achieved precisely in Type II/X for $\tan\beta = 0.96$. This only accounts for the ratio of the ditau and diphoton signals. 
\\
\\
We now consider the production rates. Neglecting the $b$ quark contribution (which is always safe for Type I and X, and safe for Type II and Y if $\tan\beta\lesssim 1$), we have
\begin{eqnarray}
    R_{gg} \sim 2.5 \xi_u^2.
\end{eqnarray}
For Type I, the branching ratios do not depend on $\tan\beta$, and we have
\begin{eqnarray}
    R_{\tau\tau} &=&0.90,\\
    R_{\gamma\gamma} &=&0.21.
\end{eqnarray}
Imposing that the signal strengths match the observed central values of $\mu_{\tau \tau}$ and $\mu_{\gamma\gamma}$, we get a very similar solution in both cases:
\begin{eqnarray}
    |\xi_u| &=& \cot\beta|\sin\theta| \sim 0.72,
\end{eqnarray}
which corresponds to $\tan\beta \sim 0.98$ for maximal mixing ($\sin\theta = \frac{1}{\sqrt{2}}$).
\\
\\
For type X, working in maximal mixing, we get
\begin{eqnarray}
    R_{\tau\tau} &=&\frac{12\tan^4\beta}{12.25+\tan^4\beta},\\
    R_{\gamma\gamma} &=& \frac{2.7}{12.25+\tan^4\beta}.
\end{eqnarray}

\noindent Once again we are able to reproduce both signals for $\tan\beta \sim 1$. The $\tau\tau$ signal also has a solution for $\tan\beta\sim 3.3$, but this does not satisfy the diphoton signal.
\\
\\
For type II and Y, in maximal mixing, the production ratio is

\begin{eqnarray}
    R_{gg} \sim \frac{1.27-0.14\tan\beta^2+0.0106\tan^4\beta}{\tan^2\beta}.
\end{eqnarray}

\noindent For type II, the decay ratios are

\begin{eqnarray}
    R_{\tau\tau} &=&\frac{1.13\tan^4\beta}{0.24+\tan^4\beta},\\
    R_{\gamma\gamma} &=& \frac{0.248}{0.24+\tan^4\beta},
\end{eqnarray}
and we can find the solutions
\begin{eqnarray}
    \tan\beta &=&0.93\quad (\gamma\gamma), \\
    \tan\beta &=& (0.53,0.88,10.7) \quad (\tau\tau).
\end{eqnarray}
We therefore find that the only good range is around $\tan\beta \sim 0.9$.
\\
\\
Finally, for type Y
\begin{eqnarray}
    R_{\tau\tau} &=&\frac{1.25}{0.37+\tan^4\beta},\\
    R_{\gamma\gamma} &=& \frac{0.28}{0.37+\tan^4\beta},
\end{eqnarray}

\noindent and we find the solution
\begin{eqnarray}
    \tan\beta &=&0.93\quad (\gamma\gamma), \\
    \tan\beta &=& 0.97 \quad (\tau\tau),
\end{eqnarray}
therefore the good range is around $\tan\beta \sim 0.95$.
\\
\\
So far we have assumed the invisible width of the pseudoscalar is zero. The pseudoscalar could instead have a small-to-moderate invisible branching ratio. However, a nonzero branching ratio would move the photon regions to lower values of $\tan\beta$, so it cannot be used to move the agreement to larger values of $\tan\beta$.
\\
\\


\noindent From this analysis, we know that there will be no viable region at large $\tan\beta$, so we focus on low $\tan\beta$. We now use the equations describing the signal strengths and impose:
\begin{eqnarray}
    0.73 < &\mu_{\tau \tau}&< 1.83,\\
    0.17 < &\mu_{\gamma \gamma}&< 0.37.
\end{eqnarray}

\begin{figure}
    \centering
    \includegraphics[width = 0.49\textwidth]{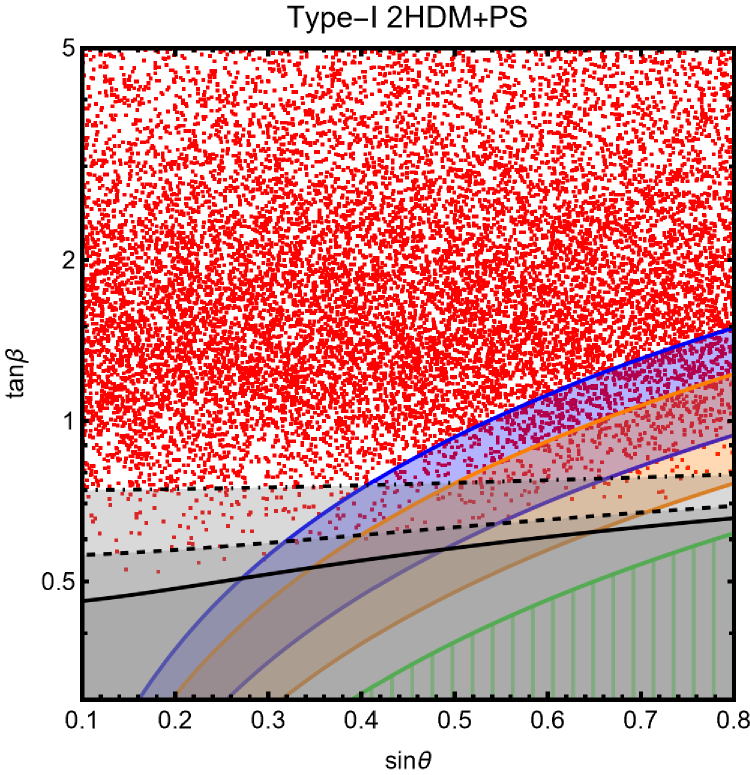}
    \includegraphics[width = 0.49\textwidth]{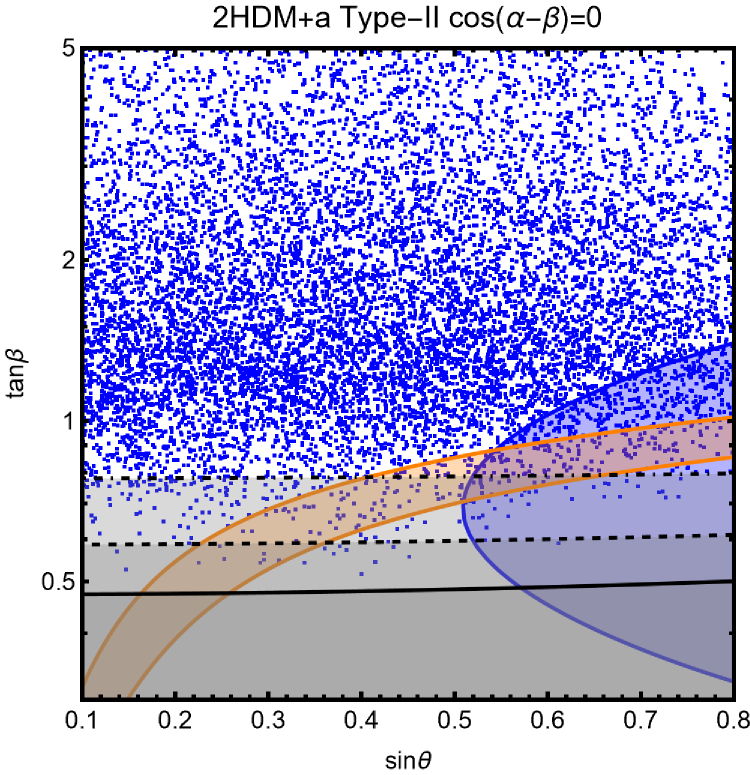}\\
    \includegraphics[width = 0.49\textwidth]{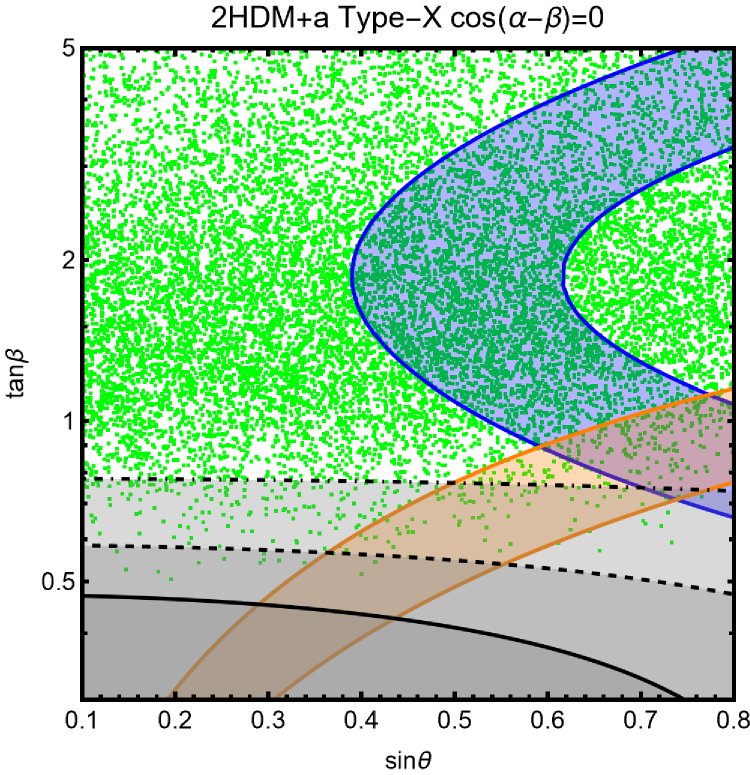}
    \includegraphics[width = 0.49\textwidth]{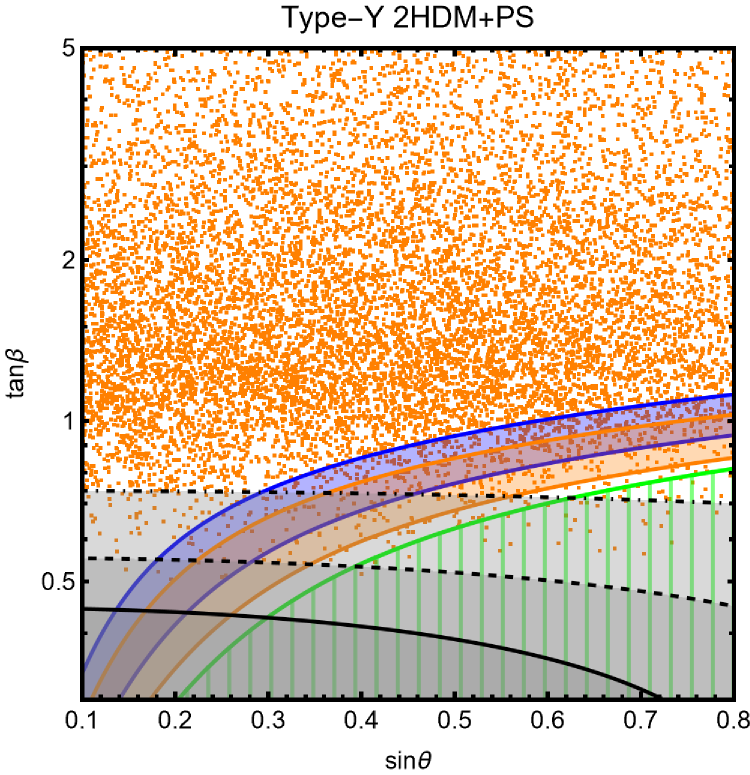}
    \caption{Regions of the 2HDM+a model compatible with CMS+ATLAS $\tau\tau$ (blue) and $\gamma\gamma$ (orange) excesses for the 4 coupling types: Type I (top left), Type II (top right), Type X (bottom left) and Type Y (bottom right). We also show regions excluded by $t\bar{t}a, a\rightarrow\tau\tau$ \citep{CMS:2024ulc} (dashed green) searches, and by flavour constraints (dark shaded regions) for different values of the heavy mediator masses: $m_A=m_H=m_{H^\pm}=1\TeV$ (dot-dashed), $1.5\TeV$ (dashed) and $2\TeV$ (solid). Colored points indicate points we found in our random scan that satisfied unitarity, BFB, and EWPT constraints. We assume the alignment limit $\cab=0$. All regions are at $2\sigma$ C.L..
    }
    \label{fig:aCOMB}
\end{figure}

\noindent These are the $2\sigma$ regions allowed by the $\tau\tau$ CMS search \citep{CMS:2022goy} and the combined \cite{CMS:2018cyk} CMS and ATLAS diphoton searches. 

We show the results of these scans in Fig. \ref{fig:aCOMB} in the $\sin\theta$ vs $\tan\beta$  plane, together with constraints from flavour physics and $t\bar{t}a$ (with $a\rightarrow \tau\tau$) limits \citep{CMS:2022arx}. The four panels of the figure show the outcomes of a parameter scan (colored points) analogous to the one performed in \cite{Arcadi:2022lpp}, done by imposing the constraints of perturbativity, unitarity and boundedness from below on the scalar potential of the 2HDM+a. The colored bands correspond to the parameter space compatible with the $\tau \tau$ (blue) and $\gamma \gamma$ (orange) excesses. The hatched green region shows the parameter space excluded by $t\bar{t}a,a\rightarrow \tau \tau$ at $95\%$ C.L..
These regions have been obtained imposing constraints on the normalised signal strengths using the values from \citep{CMS:2024ulc} obtained through \citep{Maguire:2017ypu}:

\begin{eqnarray}
    \mu_{t\bar{t}a} &=& R_{tt} R_{\tau\tau} = \xi_u^2 \frac{BR(\phi\rightarrow \tau\tau)}{BR(\phi\rightarrow \tau\tau)_{SM}} < 0.13.\label{eq:tintPS}
\end{eqnarray}
For reference we have shown, as gray regions, the exclusions from flavor bounds considering $m_A=m_H=m_{H^{\pm}}=1,1.5,2\,\mbox{TeV}$. All regions are at $2\sigma$ C.L..

All Yukawa types have compatible regions at $0.5<\tan\beta<1$. Type II and X require large mixings of $\sin\theta>0.5$ and $\sin\theta>0.6$ respectively, while Type I and Y can allow smaller mixing at lower values of $\tan\beta$.  The regions compatible with both signals are constrained mostly by the lower limits on the signals, which do not allow $\tan\beta$ to be too large as it would over-suppress the production of the pseudoscalar via gluon fusion. 
\\
\\
The $t\bar{t}a$ production cross section is very low for the pseudoscalar, so despite the updated constraints from \citep{CMS:2024ulc}, the associated limit (green dashed region) is not yet able to exclude the relevant regions in Type I and Y, while it provides no limit for Type II and X.

Limits from flavour physics depend strongly on the masses of the heavy scalars in the second Higgs doublet. We show the most constraining limit for each model Type for $m_H=m_A=m_{H^\pm}=1\TeV,1.5\TeV,2\TeV$ (dot-dashed, dashed, solid lines respectively). The black shaded regions indicate the regions excluded by the combination of flavour constraints. 

For Type I, $\epsilon_K$ and $B_s \rightarrow \mu\mu$ are the most constraining limits. The two giving similar constraints for moderate mixing angles or in the lower part of the mass range, while the latter dominates at large mixing angles and mediator masses. 

For Type X, the $\epsilon_K$ limit is the same as for Type I, but $B_s \rightarrow \mu\mu$ is suppressed at large mixing and masses. This is the reverse of the constraints for Type I: the limits are now comparable at small mixing and in the low mass range, while the $\epsilon_K$ limit dominates over the $B_s \rightarrow \mu\mu$ one for large mixing or masses. 

For Type II, the $b \rightarrow s\gamma$ constraint dominates at very low masses, but is otherwise very suppressed compared to the other ones. As in Type I and X, $\epsilon_K$ and $B_s \rightarrow \mu\mu$ once again provide the most constraining limits, with the latter giving slightly more stringent constraints for all mixing angles and in the whole mass range considered.

Finally, Type Y is similar to type X, with $\epsilon_K$ and $B_s \rightarrow \mu\mu$ giving similar constraints at small mixing angles and/or in the lower end of the mass range. $B_s \rightarrow \mu\mu$ is suppressed at large mass and/or mixing, resulting in $\epsilon_K$ being the dominant constraint in all cases.

In conclusion, in the 2HDM+a model all Yukawa types can fit the $\gamma\gamma$ and $\tau\tau$ signals. Flavour constraints can restrict the fitting regions by setting lower limits on the allowed values of $\tan\beta$ and the mixing angle $\theta$, but each Yukawa type has an allowed region at moderate to large mixing angle and for $\tan\beta\sim1$. Our numerical scan was also able to successfully populate these regions.
\\
\\
We now comment briefly on the possible DM scenarios related to this interpretation. In this section we have assumed that the new pseudoscalar has no invisible width. In Sec. \ref{sec:pheno} we found that having no invisible decay width is not a strong requirement to reproduce the signals in the case of the pseudoscalar. However, if one assumes the coupling configurations of the $\mathcal{Z}_2$ symmetric 2HDM, due to the production rates required we cannot have a large invisible branching fraction. Therefore, scenarios with $m_\chi>m_a/2\sim 47\GeV$ or $y_\chi\ll 1$ are preferred. Assuming the standard freeze-out paradigm for the DM relic density the scenario $m_\chi < m_a/2$ and $y_\chi \ll 1$ is disfavored since it would correspond to a DM annihilation cross-section sensitivity below the thermally favored value. The appropriate value of the DM annihilation cross-section can instead be easily achieved for $m_\chi > m_a/2$. However one has to account for limits from Indirect Detection, when the relic density is achieved mostly via annihilations into SM fermion pairs. Even if the relevant interactions emerge at one loop \citep{Bell:2018zra,Arcadi:2017wqi,Abe:2018emu,Ertas:2019dew}, Direct Detection constraints can be relevant in light of the very high sensitivity of present experiments. Once all the latter constraints are accounted for the most favorable scenario has $m_\chi \gtrsim 200\,\mbox{GeV}$, with the DM particle primarily annihilating into $aa$ and/or $ha$ final states.

\subsection{Interpreting the excesses in the 2HDM+s model}
\label{sec:2hdmsint}

\begin{figure}
    \centering
    \includegraphics[width = 0.49\textwidth]{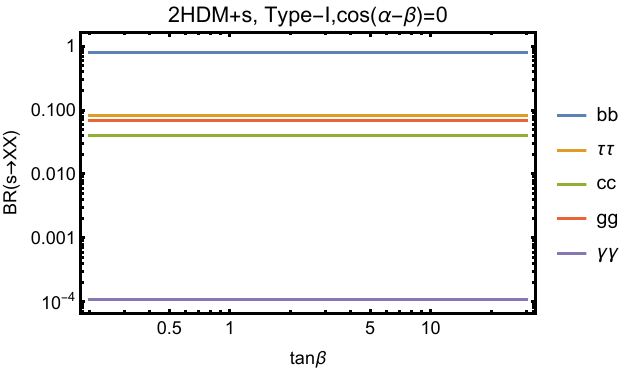}
    \includegraphics[width = 0.49\textwidth]{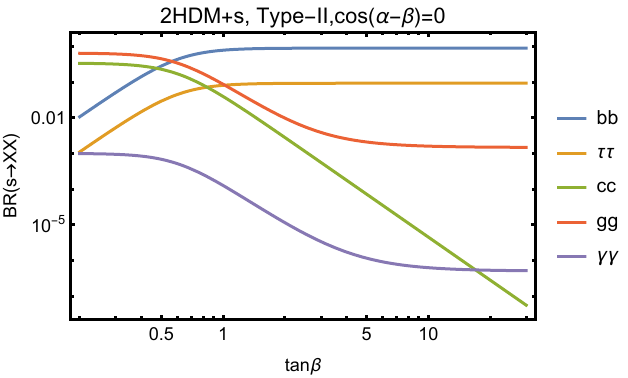}\\
    \includegraphics[width = 0.49\textwidth]{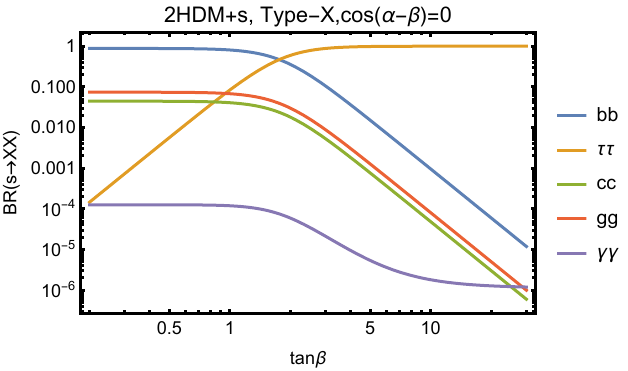}
    \includegraphics[width = 0.49\textwidth]{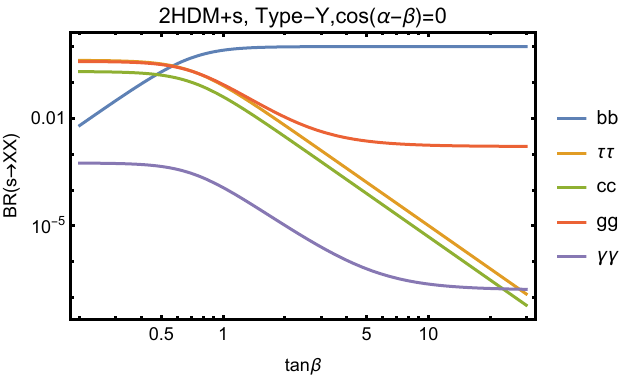}
    \caption{Decay branching ratios of the scalar $s$ particle for the different coupling configurations: Type I (top left), Type II (top right), Type X (bottom left), Type Y (bottom right). Invisible BR is assumed to be zero, and we assume the alignment limit $\cab=0$. 
    }
    \label{fig:sdecayaligned}
\end{figure}

\begin{figure}
    \centering
    \includegraphics[width = 0.49\textwidth]{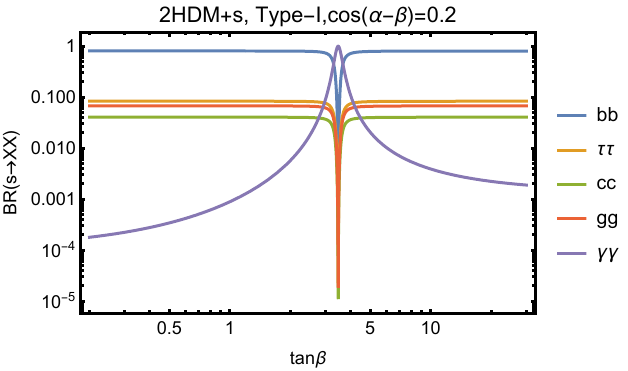}
    \includegraphics[width = 0.49\textwidth]{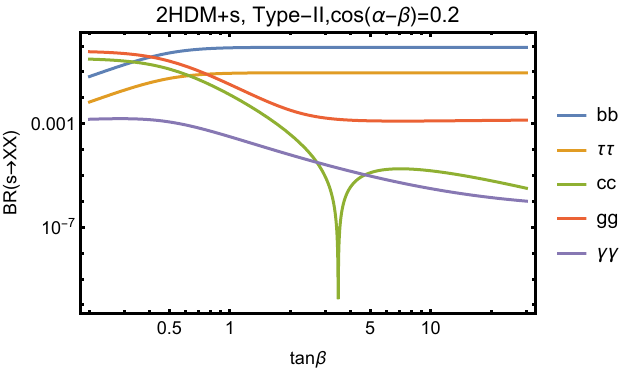}\\
    \includegraphics[width = 0.49\textwidth]{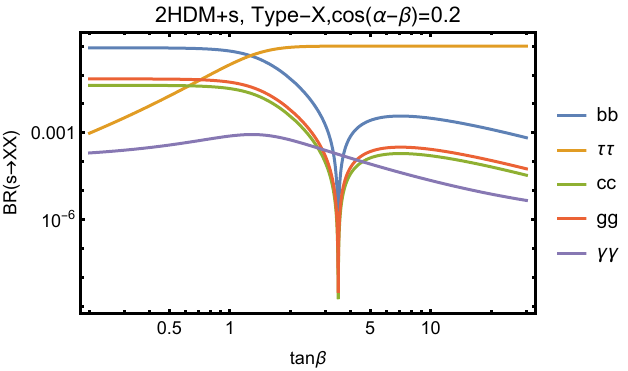}
    \includegraphics[width = 0.49\textwidth]{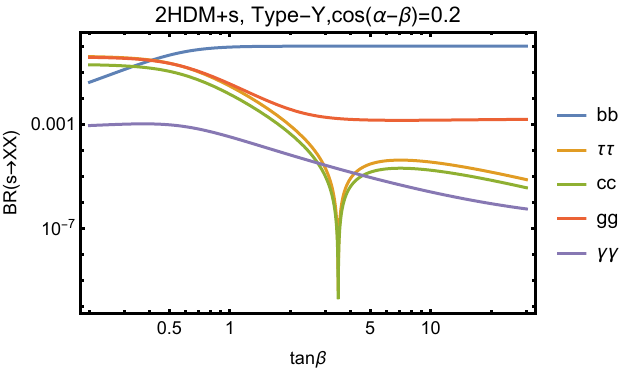}
    \caption{Decay branching ratios of the scalar $s$ particle for the different coupling configurations: Type I (top left), Type II (top right), Type X (bottom left), Type Y (bottom right). Invisible BR is assumed to be zero, and we set $\cab=0.2$. 
    }
    \label{fig:sdecayna}
\end{figure}

As with the 2HDM+a, we start our analysis for the 2HDM+s by calculating the decay branching ratios of the light scalar for this model. The results are shown in Fig. \ref{fig:sdecayaligned} for all types of 2HDM couplings and assuming the alignment limit. In Fig. \ref{fig:sdecayna} we show the widths assuming $\cos(\alpha-\beta)=0.2$ and $\sin\theta=-\frac{1}{\sqrt{2}}$. When $\cos(\alpha-\beta)$ and $\sin\theta$ have opposite signs the branching ratio to a photon pair gets positive interference effects, while if they have the same sign it gets negative interference effects. Given that the main challenge in fitting the signal is reaching the required strength for the diphoton channel, opposite-sign values of $\cos(\alpha-\beta)$ and $\sin\theta$ are preferred. We assume no invisible decay width.
\\

\noindent  We next consider the ratio $R_{\tau\gamma}$ for the scalar model. Neglecting $\tau$ and $b$ contributions in the photon loop, we get
\begin{eqnarray}
    R_{\tau\gamma} &\sim& \frac{35.5\xi_l^2}{(1.74\xi_u-7.63\xi_w)^2}.
\end{eqnarray}

\noindent Requiring it to have a value in the allowed region
\begin{eqnarray}
    2 \lesssim R_{\tau\gamma}\lesssim 10.9,
\end{eqnarray}

\noindent and substituting the couplings of Eq. \ref{eq:svvcoupling}, \ref{eq:sffcoupling}, we get the following qualitative picture: for Types I and X, allowed solutions tend to lie close to alignment at larger $\tan\beta$, regardless of the sign of $\sin\theta$. Alignment is also allowed at low $\tan\beta$, and therefore we do not get any useful information about it: as with the  pseudoscalar case, $R_{\tau\gamma}$ does not predict any value of $\tan\beta$ for Type I and X. For Type II and Y, requiring solutions at $|\cab|<0.3$ and $|\sin\theta|\ge 0.5$ (which is desirable to achieve the right production rates), forces 
\begin{eqnarray}
    \tan\beta &\lesssim& 2.3.
\end{eqnarray}

\noindent Next, we analyse production rates, starting from the $b\bar{b}$ anomaly. The production rate at LEP is
\begin{eqnarray}
    R_{ZZ} &=& g_{sVV}^2 = \cos^2(\alpha-\beta).
\end{eqnarray}

\noindent Large values of this parameter are disfavoured by Higgs physics. Therefore, to reproduce the $b\bar{b}$ signal we require a $BR_{bb}$ close to $100\%$, as in the SM. This is always the case in Type I, and is achieved in Type II and Y for $\tan\beta\gtrsim 1$ and in Type X for $\tan\beta\lesssim 1$. 
In such regions, we will require
\begin{eqnarray}
    \cos^2(\alpha-\beta) &\gtrsim& 0.06\rightarrow    |\cos(\alpha-\beta)| \gtrsim 0.24.
\end{eqnarray}
\noindent This will usually produce tension with Higgs physics constraints.

At the LHC, the production rate when $b$ quark effects can be ignored (Type I, X), and assuming VBF+VH production is negligible compared to gluon fusion, is:

\begin{eqnarray}
    R_{gg} &=& 2.47\left(0.68\cos(\alpha-\beta)+\frac{0.48\sin(\alpha-\beta)}{\tan\beta} \right).
\end{eqnarray}

\noindent Using the expressions for the branching fractions, we get 2 solutions for Type I:
\begin{eqnarray}
    \cos(\alpha-\beta) &=& 0.15, \quad \tan\beta = 0.588\\
    \cos(\alpha-\beta) &=& -0.6, \quad \tan\beta = 1.29.
\end{eqnarray}
\noindent Note that the first solution falls close to the region that reproduces the $b\bar{b}$ anomaly, but is not allowed by Higgs constraints, while the second solution does not satisfy either of the two conditions.

For Type X, we find no solutions for the central values. There are 2 regions that have some agreement, but they are both at large $\cos^2(\alpha-\beta)$. One of them is close to the region of wrong-sign Yukawa, at $\cos(\alpha-\beta)\sim0.6,\tan\beta\sim2.95$ and is close to the region favoured by the $b\bar{b}$ anomaly, while the other does not overlap well with the $b\bar{b}$ anomaly region. Neither of them fall in the region favoured by Higgs constraints.

Repeating the same procedure for Type II, we find 3 solutions at 
\begin{eqnarray}
    \cos(\alpha-\beta) &=& -0.55, \quad \tan\beta = 0.24,\\
    \cos(\alpha-\beta) &=& -0.6, \quad \tan\beta = 0.35,\\
    \cos(\alpha-\beta) &=& -0.8, \quad \tan\beta = 5.66.
\end{eqnarray}

\noindent The second solution would also be in agreement with the $b\bar{b}$ excess, however, all these solutions are excluded by Higgs constraints. Additionally, there is a region at $-0.1<\cos(\alpha-\beta)<0.4$ and low $\tan\beta$ where the signal rates are not in agreement with the central values, but are still reasonably close to them. This region also has an overlap at $\cab\sim0.35$ with the one favoured for the explanation of the $b\bar{b}$ excess, but overlaps with the region allowed by Higgs constraints only close to to the alignment limit $\cab\sim0$.

Finally, for type Y, we find 2 solutions:
\begin{eqnarray}
    \cos(\alpha-\beta) &=& 0.1, \quad \tan\beta = 0.73,\\
    \cos(\alpha-\beta) &=& -0.5, \quad \tan\beta = 1.93.
\end{eqnarray}
\noindent These solutions are close to the region that generates the required $b\bar{b}$ excess, but don't match perfectly. Additionally, only the first one is close to the region compatible with Higgs physics constraints. 
\\


\begin{figure}
    \centering
    \includegraphics[width = 0.49\textwidth]{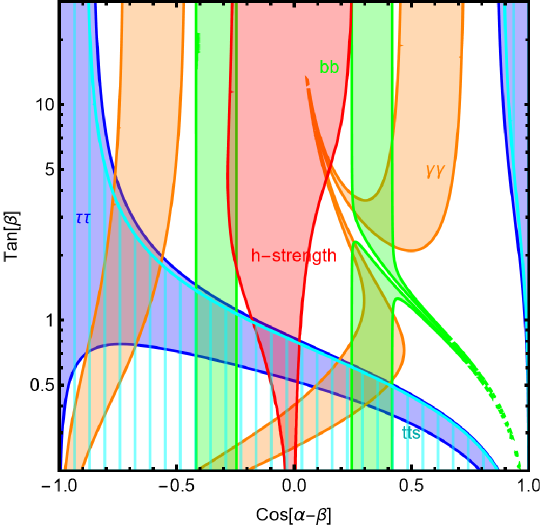}
    \includegraphics[width = 0.49\textwidth]{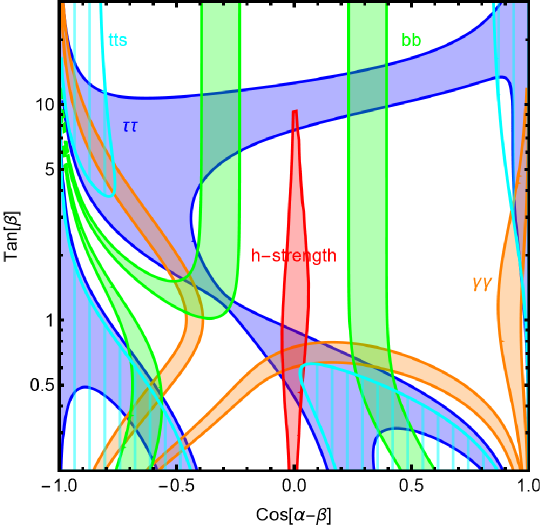}\\
    \includegraphics[width = 0.49\textwidth]{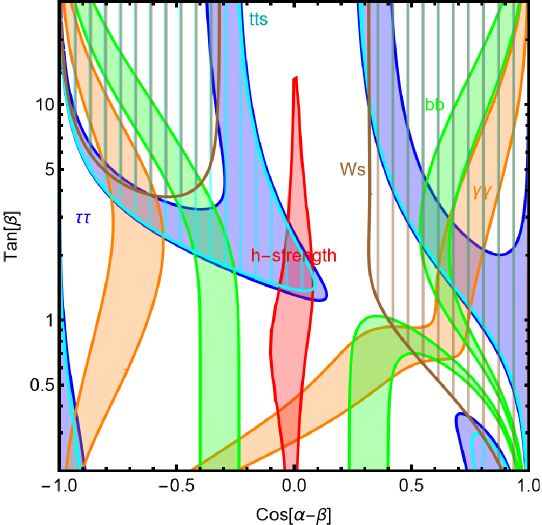}
    \includegraphics[width = 0.49\textwidth]{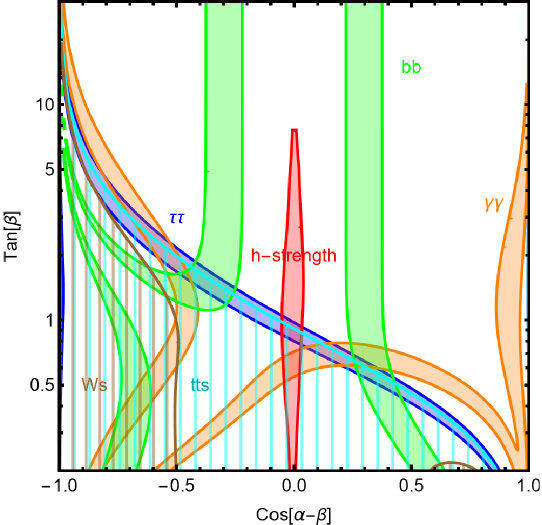}
    \caption{Regions compatible with CMS+ATLAS $\tau\tau$ (blue) and $\gamma\gamma$ (orange) excesses and the LEP $b\bar{b}$ (green) excess at $2\sigma$ C.L.. We also plot regions of compatibility at $2\sigma$ C.L. from Higgs physics (red) \citep{Bahl:2022igd}, and the regions excluded by $t\bar{t}s$ searches (dashed cyan) and $Ws$ (dashed brown) from \citep{CMS:2024ulc}. The region of compability with higgs physics has been obtained with \HiSi requiring $\chi_{hs}^2<6.18$, while all other regions are obtained by requiring the signal strengths to lie in the $2\sigma$ C.L. interval (Eq. \ref{eq:tauint},\ref{eq:gammaint},\ref{eq:bint} for $\tau\tau,\gamma\gamma,b\bar{b}$ respectively, and Eq. \ref{eq:tint},\ref{eq:wint} for $t\bar{t}s, Ws$ respectively).}
    \label{fig:sCOMB}
\end{figure}

\noindent As there will be no viable region for producing the excesses at large $\tan\beta$, we again focus on low $\tan\beta$. We impose:
\begin{eqnarray}
    0.73 < &\mu_{\tau \tau}&< 1.83, \label{eq:tauint}\\
    0.17 < &\mu_{\gamma \gamma}&< 0.37, \label{eq:gammaint}\\
    0.06 < &\mu_{bb}&<0.174. \label{eq:bint}
\end{eqnarray}

\noindent
A first summary of the results is shown in Fig. \ref{fig:sCOMB}. The four panels of the figures, corresponding to the Type-I, -II, -X and -Y scenarios, show in the $(\cos(\beta-\alpha),\tan\beta)$ plane the region satisfying the requirements on the $\tau \tau$, $\bar b b$ and $\gamma \gamma$ signal strengths stated above. The regions have been determined setting $\sin\theta=-1/\sqrt{2}$, corresponding to best-case scenario to increase the production rate. The plots also highlight, in red, the regions corresponding to coupling of the 125 GeV Higgs compatible, within $2\sigma$, with the experimental constraints \citep{ATLAS:2015ciy,ATLAS:2021vrm}. These regions have been determined via the \HiTo package \citep{Bahl:2022igd}. Finally, the hatched cyan and brown region are ruled out, always at the $95\,\%$ C.L., by searches of $t\bar ts$ and $Ws$ events \citep{CMS:2024ulc}. These regions have been obtained imposing constraints on the normalised signal strengths using the values from \citep{CMS:2024ulc} obtained through \citep{Maguire:2017ypu}:

\begin{eqnarray}
    \mu_{t\bar{t}s} &=& R_{tt} R_{\tau\tau} = \xi_u^2 \frac{BR(\phi\rightarrow \tau\tau)}{BR(\phi\rightarrow \tau\tau)_{SM}} < 0.065,\label{eq:tint}\\
    \mu_{Ws} &=& R_{ZZ}R_{\tau\tau} = \cab^2 \frac{BR(\phi\rightarrow \tau\tau)}{BR(\phi\rightarrow \tau\tau)_{SM}} < 0.1.\label{eq:wint}
\end{eqnarray}

In summary, one would apparently obtain a viable fit of the 95 GeV excesses where there is overlap between the blue/green/orange/red regions but outside the hatched regions. 

Note that such constraints are weaker than in the 2HDM, because the couplings inherited from the second doublet are suppressed by $|\sin\theta|=1/\sqrt{2}$. 

We can identify in the figures all the regions that we found analytically for the various Yukawa types. As expected, all Yukawa types have some small regions that are compatible with all excesses, but none of these regions survive current Higgs or collider constraints. All Yukawa types, with the exception of Type X, have a region at $\tan\beta\sim 0.6$ where there is a near-overlap of all 3 regions, but does not overlap with the region allowed by Higgs physics constraints for any type, and overlaps with the one excluded at $2\sigma$ C.L. from $t\bar{t}s$ for Type I and Y, where the $t\bar{t}s$ exclusion region closely tracks the region where the $\tau\tau$ excess is produced. For Type II, there is a near-overlap of the intersection and the exclusion from $t\bar{t}s$.

Type I and X don't seem able to accommodate the $\tau\tau$ signal at all, as the regions required to reproduce that signal are nearly identical to the ones excluded by $t\bar{t}s$. Type I has some very small overlaps between the Higgs physics allowed region, the region allowed by $t\bar t s$ one of the $\gamma\gamma$ or $b\bar{b}$ allowed regions separately (the $\gamma\gamma$ requires fine-tuning), being able to accommodate at most 1 of the 3 excesses. For Type X, the region allowed by Higgs physics has some overlap only with the $\gamma\gamma$ region. Type II seems to be able to potentially accommodate $\gamma\gamma$ and $\tau\tau$ in the alignment region at $\tan\beta\sim0.7$. Finally, Type Y does not seem to have any region allowed by Higgs constraints and $t\bar{t}s$ that can reproduce any of the signals.

Finally, constraints from flavour physics are not shown in Fig. \ref{fig:sCOMB}, but all models behave very similarly, with $B_s \rightarrow \mu^+\mu^-$ always dominating flavour bounds, and setting a lower limit on $\tan\beta\gtrsim 0.74,0.55,0.45$ for $m_A=m_H=m_{H^\pm}=1\TeV,1.5\TeV,2\TeV$ respectively.
\\

\begin{figure}
    \centering
    \includegraphics[width = 0.49\textwidth]{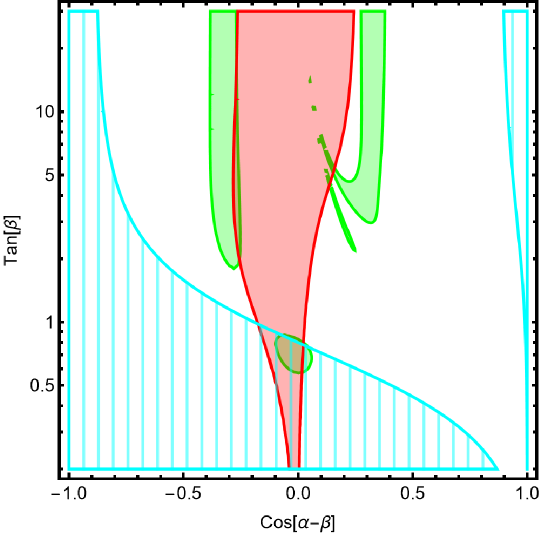}
    \includegraphics[width = 0.49\textwidth]{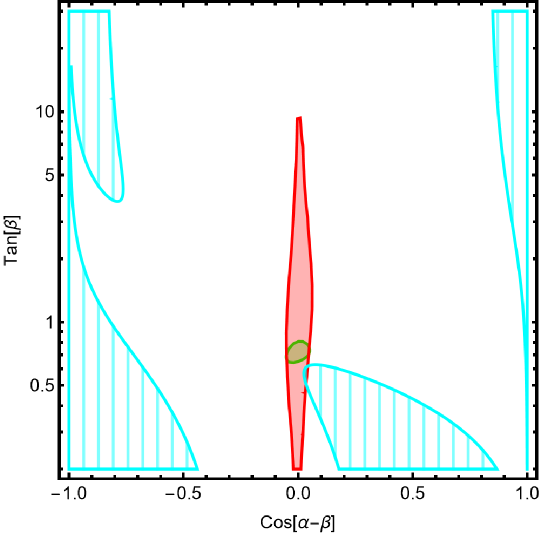}\\
    \includegraphics[width = 0.49\textwidth]{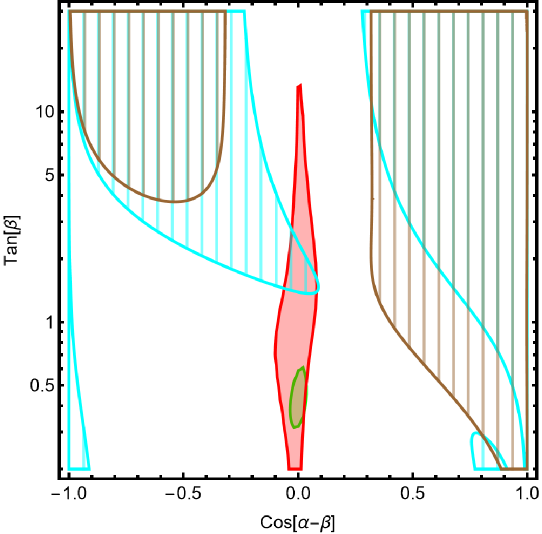}
    \includegraphics[width = 0.49\textwidth]{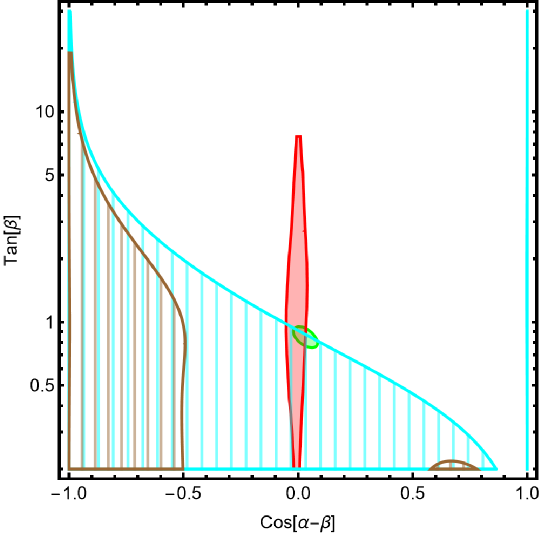}
    \caption{
    In green, the $2\sigma$ C.L. regions obtained by the combined $\chi_{st}^2$ analysis done with \HiSi and the $\chi_{95}^2$ analysis for the $95\GeV$ particle outlines in the text. In red, the $2\sigma$ C.L. regions obtained by the $\chi_{hs}^2$ analysis only. We report, for reference, the $2\sigma$ exclusion regions for the $t\bar{t}s$ (dashed cyan) and $Ws$ (dashed brown) searches from \citep{CMS:2024ulc}.
    \label{fig:sCOMBchi}
    }
\end{figure}

\noindent To make a more rigorous assessment, from a statistical perspective, of our findings, we have introduced a $\chi^2$ test. We hence define:
\begin{equation}
    \chi^2=\chi_{hs}^2+\chi_{95}^2
\end{equation}
with $\chi_{hs}^2$ being the $\chi^2$ provided by \HiSi \citep{Bahl:2022igd,Bechtle:2020uwn} and accounting for all the relevant constraints. $\chi_{95}^2$ combines instead the signal rates associated to the excesses:
\begin{eqnarray}
    \chi_{95}^2=\sum_{i} \left(\frac{R_i-R_i^{exp}}{\Delta R_i^{exp}}\right)^2.
\end{eqnarray}
The entries of the latter expression are listed in Tab. \ref{table:chi95}.

\begin{table}
\renewcommand{\arraystretch}{1.17}
\begin{center}
\begin{tabular}{|c|c|c|c|c|c|}
\hline
 & $R_{\mu\mu}$ & $R_{\gamma\gamma}$ &  $R_{b\bar{b}}$ & $R_{t\bar{t}s}$ & $R_{Ws}$ \\ \hline \hline
$R_i^{exp}$ & $1.225$ & $0.27$ & $0.117$ & $0$ & $0$  \\ \hline
$\Delta R_i^{exp}$ & $[-0.24,+0.31]$ & $0.10$ & $0.057$ & $0.033$ & $0.05$  \\ \hline\hline
\end{tabular}
\vspace*{2mm}
\caption{Values used in the $\chi_{95}^2$ analysis. $R_{t\bar{t}s}$ and $R_{Ws}$ are from \citep{CMS:2024ulc}.
}
\label{table:chi95}
\end{center}
\vspace*{-6mm}
\end{table}

The results of the $\chi^2$ analysis are shown in Fig. \ref{fig:sCOMBchi}, again in the $\cos(\beta-\alpha,\tan\beta)$ plane. This time the regions marked in green correspond to the condition  $\chi^2-\chi_{MIN}^2<6.18$, namely a combined fit of the excesses at the $2\sigma$ C.L.\footnote{In our analysis, we report exclusion and compatibility regions corresponding to a $2\sigma$ C.L. for a normal distribution, which is approximately $95\%$ C.L. for a generic distribution.}. For comparison, the red regions corresponding to $\chi_{hs}^2-\chi_{hs,MIN}^2<6.18$, i.e. the fit of the observables related to the 125 GeV, as given by \HiSi, and the excluded regions by BSM searchers, as given by \HiBo, have been shown as well.

By comparing Fig. \ref{fig:sCOMBchi} with Fig. \ref{fig:sCOMB} we see that there are 3 very narrow regions, in the Type-I case, satisfying the $\chi^2$ conditions. The reason can be understood by comparing  Fig. \ref{fig:sCOMBchi} with Fig. \ref{fig:sCOMB}. Indeed, in the Type-I, the different signals, namely $\tau \tau$, $b \bar b$ and $\gamma \gamma$ tend to prefer different regions of the parameter space. 
Furthermore, as evidenced by Fig. \ref{fig:sCOMBchi}, the 95 GeV excesses are in evident tension with the experimental constraints from Higgs signal strength and $t\bar{t}s$ and $Ws$ searches. For what the Type-II is concerned, we notice a single green region, close to the alignment limit, which is in good agreement with the Higgs signal strength and evades the exclusion from BSM searches. However, a closer inspection of $\chi^2$ and a comparison with Fig. \ref{fig:sCOMB} suggest a poor agreement with the $b\bar b$ excess. A similar outcome has been achieved also for the type-X configuration but with a poor fit of both the $b\bar b$ and the $\tau \tau$ signals. Moving finally to the type-Y, the $\chi^2$ test identifies again a single region for $\tan\beta \simeq 0.8$. This region corresponds to a good fit of the $\tau \tau$ signal, a marginal fit of the $\gamma \gamma$ while it is substantially incompatible with the $\bar b b$ excess. Fig. \ref{fig:sCOMBchi} also evidences a strong tension with $t \bar {t} s$ constraints.
In summary we find that for the 2HDM+s, no one of the considered Yukawa configurations is capable of accounting for successfully all the, $b \bar b$, $\tau \tau$ and $\gamma \gamma$, excesses and, at the same time, evade the complementary collider constraints. Relaxing the requirement of the $\bar b b$ excess, the Type-II scenario provides a good fit of the two LHC signals.
\\
\\
The reasons why the 2HDM+s is mostly unsuccessful in reproducing the signal under scrutiny are the following. First of all we have problems intrinsic to the model. Indeed, reproducing the $b\bar{b}$ signal requires disalignment, in clear contrast with Higgs signal strength constraints, pointing towards near-alignment. On a similar footing, achieving production rates and decay branching fraction of the resonance to photons compatible with the $\gamma \gamma$ excess requires deviation from the alignment limit and/or low values of $\tan\beta$ (to enhance the coupling with the top quarks). Additionally, the recent update by CMS \citep{CMS:2024ulc} of searches of $t \bar t s$ events, and the constraints derived from this result, present a final reason why the 2HDM+s cannot reproduce the all three signals.

\section{Conclusions}
\label{sec:conclusion}

In this paper we have considered possible signal excesses that have arisen in data from LHC ($\tau\tau$ \citep{CMS:2022goy} and $\gamma\gamma$ \citep{CMS:2023yay}) and LEP ($b\bar{b}$ \citep{LEPWorkingGroupforHiggsbosonsearches:2003ing,Azatov:2012bz,Cao:2016uwt}), all at similar mass values of $\sim95\GeV$. We have tried to interpret these excesses in terms of the production and decay of a new scalar or pseudoscalar particle. We have first performed a generic analysis, without assuming a specific model and considering the Natural Flavour Conservation scenario. This constrains the couplings of the Standard Model fermions to a new (pseudo)scalar. Assuming that there are no new charged particles of mass $\sim\mathcal{O}(100\GeV)$, these constraints set a relation between the fermion and photon coupling, and therefore can be used to constrain the interpretation in terms of a new scalar or pseudoscalar. We find that the interpretation in terms of a new pseudoscalar particle can accommodate the $\tau\tau$ and $\gamma\gamma$ signal, while it cannot accommodate the $b\bar{b}$ excess at LEP as there is no efficient mechanism for a pseudoscalar at LEP to reproduce the anomaly. The LHC production rates for a pseudoscalar through gluon fusion are however more than enough to accommodate the $\tau\tau$ and $\gamma\gamma$ signals, while also allowing a nonzero invisible branching fraction. In the scalar case we instead find that, while it might be possible to accommodate all 3 signals, it becomes challenging to reproduce the observed rates. This is due to the lower production rates for a scalar at LHC, and the fact that the production at LEP requires some mixing of the new scalar with the Standard Model Higgs boson, which is tightly constrained by Higgs physics.
\\
\\
We have then moved on to perform an analysis of these excesses in terms of a $\mathbb{Z}_2$-symmetric 2HDM with an additional scalar or pseudoscalar singlet. In these models, the additional Higgs doublet generates four additional fields (a scalar $H$, pseudoscalar $A$, and charged scalars $H^{\pm}$) after EWSB. The additional (pseudo)scalar singlet $(a)s$ mixes with these fields, inheriting couplings to SM fermions despite being gauge singlets. 

The four possible Yukawa types of the 2HDM impose additional relations between the couplings of the new scalar to the SM fermions, leaving less freedom to accommodate the new signals. 

\noindent
We have applied to both models the usual theoretical (Unitarity and Boundedness from Below) and experimental (flavour, electroweak precision tests (EWPT), Higgs physics and collider) constraints. Unitarity and EWPT are known to limit the mass splittings between the new scalar particles, while Higgs physics constraints usually restrict the 2HDM parameter space to regions close to the alignment limit, with the possible exception of the Type-I Yukawa configuration. Limits from flavour physics are not very sensitive to the mass of the new neutral scalars, including the particle at $95\GeV$, but instead mostly affect the mass of the new additional charged scalar and, most importantly, $\tan\beta$. We assume that all other scalars are much heavier and quasi-degenerate, so that flavour physics constraints can be directly translated into lower limits on the value of $\tan\beta$ that range from $\tan\beta\gtrsim0.5$ to $\tan\beta\gtrsim0.8$, depending on the model, the mass of the heavy scalars (we consider values from $1\TeV$ to $2\TeV$), and the specific Yukawa configuration.

\noindent
Direct production of the new (pseudo)scalar at colliders can happen via gluon fusion or in association with a top quark pair $t\bar{t}$. Limits from the production in association with $t\bar{t}$ are then crucial to assess the viability of the models.

The largest branching fractions of the new scalar are expected to be to $b\bar{b}$ and $\tau\tau$. However, a $b\bar{b}$ final state would not be observable at LHC due to the large QCD background, and so for this channel we can only rely on LEP data. Alternatively, one can consider a final state where the new scalar is produced in association with another SM boson (such as $Z$ or $h_{125}$), or double production and decay to a 4 bottom final state, both of which can originate from resonant production and the subsequent decay of one of the heavier scalars $H,A$. 

In addition to the $b\bar{b}$ and $\tau\tau$ final states, the $\gamma\gamma$ final state could also be detected, and these three are in fact the ones where we consider excess signals. Additional collider constraints can come from other cascade decay searches, where one of the heavy scalars is resonantly produced and decays to SM particles and/or the light (pseudo)scalar. Searches of this type have been considered and can usually constrain the mass of the additional scalar to be above $\sim1\TeV$.

Our main findings can be summarized as follows. Neither of the models considered can account for the $b\bar b$, $\tau \tau$ and $\gamma \gamma$ signals at the same time, while being compatible with the complementary experimental constraints.
The 2HDM+a is intrinsically incompatible with the $b \bar b$ as CP invariance forbids the needed couplings between the pseudoscalar resonance and the gauge bosons. If only the LHC excesses are considered instead, all the four Yukawa configurations provide a good fit in the alignment limit and are moderately affected by flavor constraints as well as $t \bar t a$ searches, thanks to the reduced production rate in the 2HDM+a. Types I and Y have regions that allow moderate mixing of the pseudoscalars, while Types II and X require larger values of mixing angles ($\sin\theta\gtrsim 0.5$ and $\sin\theta\gtrsim 0.6$ respectively).
The case of 2HDM+s is more complex. In order to account for the $\bar b b$ and $\gamma \gamma$ signal strengths, deviations from the alignment limit are required, in strong tension with experimental evidence. Furthermore, a strong impact on the parameter space is due to limits from $\bar t t s$ searches. Nevertheless, a possible fit of the $\gamma \gamma$ and $\tau \tau$ signals is achieved for the Type-II Yukawa configuration.

A possible workaround to these limitations was considered in \citep{Azevedo:2023zkg}, in the context of a pure 2HDM where a CP-breaking scenario is considered. In such a model, one could have a light spin-0 particle at $95\GeV$ that has mixed scalar/pseudoscalar couplings to fermions and non-zero couplings to gauge bosons. This scenario would likely alleviate the second point of tension by allowing larger loop-level couplings to gluons and photons while also allowing a fit for the $b\bar{b}$ signal. The CP-breaking scenario would also turn $h_{125}$ into a mixed CP state, but this would likely still be allowed by current data as the CP phase of the Higgs couplings to fermions is not well-constrained \citep{Haisch:2016gry,Mb:2022rxu,Azevedo:2022jnd,CMS:2022dbt}. However, the main issue in such a scenario is that the presence of light CP-breaking scalars at $95\GeV$ and $125\GeV$ can strongly conflict with Electron Dipole Moment (EDM) constraints \citep{Abe:2013qla}. The 2HDM+s model could naturally induce some cancellation of the EDM amplitudes due to the scalar mixing structure it features and the fact that the two light scalars would have similar masses ($95\GeV$ for the new scalar and $125\GeV$ for the Higgs boson). However, the model would still need a considerable amount of tuning to avoid such constraints. As such, we choose not to quantitatively investigate this scenario, leaving it for future work.

It remains to be seen if these excesses are connected to a real anomaly or are just statistical fluctuations. The experimental excesses are mainly driven by the results of the CMS collaboration, and new results from LHC Run 3 will hopefully shed more light on the origin of these excesses.


\section*{Acknowledgements}
GB was supported by the Australian Government through the Australian Research Council Centre of Excellence for Dark Matter Particle Physics (CDM, CE200100008). NK was supported by an Australian Government Research Training Program Scholarship. The authors thank Jo\~{a}o Seabra, Jo\~{a}o Paulo Pinheiro, Fabio Risitano and Alessandro Pilloni for useful conversations.

\newpage

\appendix

\section{Decay Widths}
\label{sec:decay}

The partial decay widths to $f\bar{f}$ are: 

\begin{eqnarray}
\Gamma(S\rightarrow ff) &=& \frac{N_c g_{sff}^2 m_f^2(M) M}{8\pi v^2}\left(1-\frac{4m_f^2}{M^2}\right)^{3/2}(1+\delta_s),\\
\Gamma(P\rightarrow ff) &=& \frac{N_c g_{pff}^2 m_f^2(M) M}{8\pi v^2}\left(1-\frac{4m_f^2}{M^2}\right)^{1/2}(1+\delta_p),
\end{eqnarray}
where $\delta_{s,p}$ are the QCD and EW corrections \citep{Choi:2021nql} and depend on the fermion/quark type. The partial decay widths to $gg$ and $\gamma\gamma$ are:
\begin{eqnarray}
\Gamma(S\rightarrow gg) &=& \frac{ \alpha_s^2 M^3}{32\pi^3v^2} |\sum_f g_{sff} F_S\left(\frac{M^2}{4m_f^2}\right)|^2(1+\delta_s),\\
\Gamma(P\rightarrow gg) &=& \frac{ \alpha_s^2 M^3}{32\pi^3 v^2} |\sum_f g_{pff} F_P\left(\frac{M^2}{4m_f^2}\right)|^2(1+\delta_p),\\
\Gamma(S\rightarrow \gamma\gamma) &=& \frac{ \alpha_s^2 M^3}{32\pi^3v^2} |\sum_f 2N_c^f Q_f^2 g_{sff} F_S\left(\frac{M^2}{4m_f^2}\right)-g_{sWW}F_W\left(\frac{M^2}{4m_f^2}\right)|^2(1+\delta_s),\\
\Gamma(P\rightarrow \gamma\gamma) &=& \frac{ \alpha_s^2 M^3}{32\pi^3 v^2} |\sum_f 2N_c^f Q_f^2 g_{pff} F_P\left(\frac{M^2}{4m_f^2}\right)|^2(1+\delta_p),
\end{eqnarray}
where
\begin{eqnarray}
F_S(x) &=& x^{-1} (1+(1-x^{-1})f(x)),\\
F_P(x) &=&x^{-1}f(x),\\
F_w(x) &=& 2+3x^{-1}+3x^{-1}(2-x^{-1})f(x),\\
f(x) &=& \arcsin^2(\sqrt{x}),\quad x\le 1,\\
f(x) &=& -\frac{1}{4}(\log\frac{\sqrt{x}+\sqrt{x-1}}{\sqrt{x}-\sqrt{x-1}} -i\pi)^2, \quad x>1,
\end{eqnarray}
and $\delta_{s,p}$ are the QCD and EW corrections \citep{Choi:2021nql}, which depend on the channel ($gg$ or $\gamma\gamma$). $g_{sff},g_{pff}$ are the ratios between the quark coupling to the spin-0 particle and the SM Yukawa couplings, and $g_{sWW}$ is the ratio between the coupling of $s$ to a $W$ pair and the same coupling of the SM Higgs.
\\
\\
For the 3-body decay $h_{125}\rightarrow af\bar{f}$, we have:

\begin{eqnarray}
\Gamma(h_{125} \rightarrow a \bar f f)=\frac{N_c^f |g_{A^0ff}|^2}{128\pi^3}\frac{m_f^2}{v^2}\frac{|g_{haa}|^2}{m_h} g\left(\frac{4 m_a^2}{m_h^2}\right) \sin^2 \theta,
\end{eqnarray}
where
\begin{eqnarray}
g(x)=\frac{1}{8}(x-4)\left[4-\log\left(\frac{x}{4}\right)\right]-\frac{5x-4}{4 \sqrt{x-1}}\left[\arctan\left(\frac{x-2}{2\sqrt{x-1}}\right)-\arctan\left(\frac{1}{\sqrt{x-1}}\right)\right],
\end{eqnarray}

\noindent and the coupling $g_{haa}$ is:
\begin{eqnarray}
    g_{haa} = -\frac{1}{v}&\left(  \left(-2 m_a^2+m_h^2-2 m_H^2+4
   m_{H^\pm}^2-2 \lambda_3 v^2\right)\sin ^2\theta\right.\nonumber\\
   &-\left.2 v^2 
   \left(\lambda_{1P} \cos ^2\beta +\lambda_{2P} \sin ^2\beta
   \right)\cos^2\theta\right).
\end{eqnarray}

\label{Bibliography}


\bibliography{Bibliography} 

\end{document}